\documentclass[sigplan,screen]{acmart}
\AtBeginDocument{%
  }

\usepackage{xspace}
\usepackage[most]{tcolorbox}
\tcbuselibrary{breakable}
\usepackage{soul,color}
\usepackage{multirow}
\usepackage{makecell}
\usepackage{pifont}
\usepackage{marvosym}
\usepackage[table,xcdraw]{xcolor}
\usepackage{booktabs}       
\usepackage{threeparttable} 
\usepackage{enumitem}
\setlist[itemize]{topsep=0pt,left=5pt}

\newcommand{\ouragent}{TermiAgent\xspace} 

\newcommand{\songbench}{TermiBench\xspace} 

\newcommand{\memorymodule}{\textit{Memory Module} } 
\newcommand{\arsenalmodule}{Arsenal Module\xspace}
\newcommand{\executormodule}{Executor Module\xspace}
\newcommand{\reasonermodule}{Reasoner Module\xspace}
\newcommand{\assistantmodule}{Assistant Module\xspace}

\begin{document}

\title{Shell or Nothing: Real-World Benchmarks and Memory-Activated Agents for Automated Penetration Testing}

\author{Wuyuao Mai}
\authornote{Both authors contributed equally to this research.}
\affiliation{%
  \institution{Fudan University}
  \city{Shanghai}
  \country{China}
}
\email{maiwuyuao20@fudan.edu.cn}

\author{Geng Hong}
\authornotemark[1]
\affiliation{%
  \institution{Fudan University}
  \city{Shanghai}
  \country{China}
}
\email{ghong@fudan.edu.cn}

\author{Qi Liu}
\affiliation{%
  \institution{Fudan University}
  \city{Shanghai}
  \country{China}
}
\email{qiliu25@m.fudan.edu.cn}

\author{Jinsong Chen}
\affiliation{%
  \institution{Fudan University}
  \city{Shanghai}
  \country{China}
}
\email{jschen23@m.fudan.edu.cn}

\author{Jiarun Dai}
\affiliation{%
  \institution{Fudan University}
  \city{Shanghai}
  \country{China}
}
\email{jrdai@fudan.edu.cn}

\author{Xudong Pan}
\affiliation{%
  \institution{Fudan University}
  \institution{Shanghai Innovation Institute}
  \city{Shanghai}
  \country{China}
}
\email{xdpan@fudan.edu.cn}

\author{Yuan Zhang}
\affiliation{%
  \institution{Fudan University}
  \city{Shanghai}
  \country{China}
}
\email{yuanxzhang@fudan.edu.cn}

\author{Min Yang\textsuperscript{\Letter}}
\affiliation{%
  \institution{Fudan University}
  \city{Shanghai}
  \country{China}
}
\email{m_yang@fudan.edu.cn}

\renewcommand{\shortauthors}{Wuyuao Mai, et al.}

\begin{abstract}

Penetration testing is critical for identifying and mitigating security vulnerabilities, yet traditional approaches remain expensive, time-consuming, and dependent on expert human labor. Recent work has explored AI-driven pentesting agents, but their evaluation relies on oversimplified capture-the-flag (CTF) settings that embed prior knowledge and reduce complexity, leading to performance estimates far from real-world practice.
We close this gap by introducing the first real-world, agent-oriented pentesting benchmark, \songbench, which shifts the goal from “flag finding” to achieving full system control. The benchmark spans 510 hosts across 25 services and 30 CVEs, with realistic environments that require autonomous reconnaissance, discrimination between benign and exploitable services, and robust exploit execution. Using this benchmark, we find that existing systems can hardly obtain system shells under realistic conditions.

To address these challenges, we propose \ouragent, a multi-agent penetration testing framework. \ouragent mitigates long-context forgetting with a Located Memory Activation mechanism and builds a reliable exploit arsenal via structured code understanding rather than naïve retrieval. In evaluations, our work outperforms state-of-the-art agents—exhi-biting stronger penetration testing capability, reducing execution time and financial cost, and demonstrating practicality even on laptop-scale deployments.
Our work delivers both the first open-source benchmark for real-world autonomous pentesting and a novel agent framework that establishes a milestone for AI-driven penetration testing.

\end{abstract}




\maketitle

\section{Introduction}

Penetration testing (pentesting) is a cornerstone of modern cybersecurity, enabling organizations to proactively identify and mitigate exploitable vulnerabilities before adversaries can capitalize on them. By simulating realistic cyberattacks, pentesting not only assesses the robustness of deployed defenses but also informs remediation strategies and compliance with security standards \cite{ref1,NIST-SP800-115}. Traditional pentesting, however, is resource-intensive—typically costing \$10,000–\$20,000 per engagement \cite{ref3,ref4}—and demands highly skilled human experts capable of navigating complex, heterogeneous systems under tight time constraints. These limitations have fueled growing interest in automation, particularly through AI-assisted cybersecurity testing.

Recent advances in large language models (LLMs) and autonomous agents have inspired new approaches to AI-assisted security testing, ranging from vulnerability discovery and patch generation (e.g., DARPA’s AI Cyber Challenge) \cite{kapko2025darpa} to AI-driven penetration testing frameworks such as PentestGPT and VulnBot \cite{kong2025vulnbotautonomouspenetrationtesting,299699,shen2025pentestagentincorporatingllmagents}. While these prototypes demonstrate the potential of LLM-based agents to accelerate parts of the pentesting workflow, their evaluations remain tethered to oversimplified, CTF-style environments. These settings often embed a priori knowledge—such as leaked passwords, predefined exploit entry-points that human red teams would not have in practice, and they set up an unrealistic testing environment that each target only hosts exactly one vulnerable service. As a result, these simplified benchmarks may overestimate real-world agent pentest performance.

In contrast, real-world penetration testing unfolds in a far more challenging operational landscape. Red-teams often begin with nothing more than network access, requiring them to perform reconnaissance and enumeration under uncertainty while distinguishing benign background services from the actual attack surface of the targets. Achieving full system compromise—culminating in an interactive shell—necessitates the integration of diverse skills, tools, and reasoning under dynamic and incomplete information. 
Current benchmarks for autonomous penetration testing agents, structured as CTF-style exercises and providing prior hints, are insufficient for evaluating agent performance in real-world scenarios.

To close these gaps, we present \songbench, the first benchmark for real-world, fine-grained, and agent-oriented penetration testing evaluation.
First, our benchmark shifts the final objective from “flag-finding” to achieving system control, specifically obtaining a system shell. To more faithfully replicate real-world conditions, the benchmark omits unnecessary prior information—such as entry points and predefined exploit paths—thereby requiring agents to conduct reconnaissance autonomously. In addition, we configure target hosts with common applications, including web servers and database systems, to introduce “noise” that compels agents to distinguish between exploitable and non-exploitable services. Finally, the benchmark comprises 510 hosts embedding 30 CVEs across 25 distinct services over a ten-year span. Each host is configured with up to seven vulnerability-free services alongside one vulnerable service dated between 2015 and 2025.

With this benchmark, we find that previous work fails to acquire the system shell of target hosts. Agents were either lost in vast amounts of exploratory information that accumulated in complex real-world penetration testing scenarios, or could not penetrate the target service without ready-to-use penetration testing agents' exploits.

In this paper, we propose \ouragent, a multi-agent framework tailored for real-world penetration testing.
To address the challenge of long-context forgetting in penetration testing, we introduce a Located Memory Activation approach. When predicting its next action, the agent automatically activates all relevant memories required for decision-making, reflecting the characteristics of real-world penetration testing tasks. To build an up-to-date and ready-to-use exploit arsenal, we formulate exploit integration as a structured code-understanding problem rather than a simple retrieval-and-execution task. Unlike naive methods that merely fetch public PoC repositories and attempt direct execution, our approach ensures robust and reliable exploit utilization.
As illustrated in Figure \ref{fig:overview}, \ouragent \ comprises the \reasonermodule, the \assistantmodule, the \executormodule, the \memorymodule and the \arsenalmodule, which collaboratively decompose penetration testing targets into multi-step sub-tasks, progressively achieving the final objectives in a perception–action loop.

To evaluate the performance of \ouragent, we compare it against state-of-the-art penetration testing agents. Results demonstrate that \ouragent solves approximately 1.7 times as many CTF challenges and over 8 times as many real-world penetration testing tasks as state-of-the-art agents.
In terms of efficiency, \ouragent achieves lower time and financial costs, requiring less than one-fifth of the execution time and only a tenth of the financial cost in real-world scenarios. Furthermore, our arsenal covers 1.8 times more RCE CVEs than Metasploit.

In summary, this papers make the following contributions:
\begin{itemize}
\item We reveal the gap between real-world penetration testing requirements and existing LLM-based approaches, showing that current benchmarks overestimate real-world agent penetration performance.
\item We construct and open-source the first real-world, fine-grained, and agent-oriented penetration benchmark that replicates real-world conditions for agent-oriented penetration testing.
\item We design and implement \ouragent, a penetration testing framework that achieves significantly better performance than prior work.
\item We demonstrate that even a laptop-scale deployment can effectively support end-to-end automated penetration testing, highlighting the practicality of our approach.
\end{itemize}
\section{Background and Preliminary Study}\label{sec:background}

\begin{figure*}[!ht]
    \centering
    \includegraphics[width=0.9\linewidth]{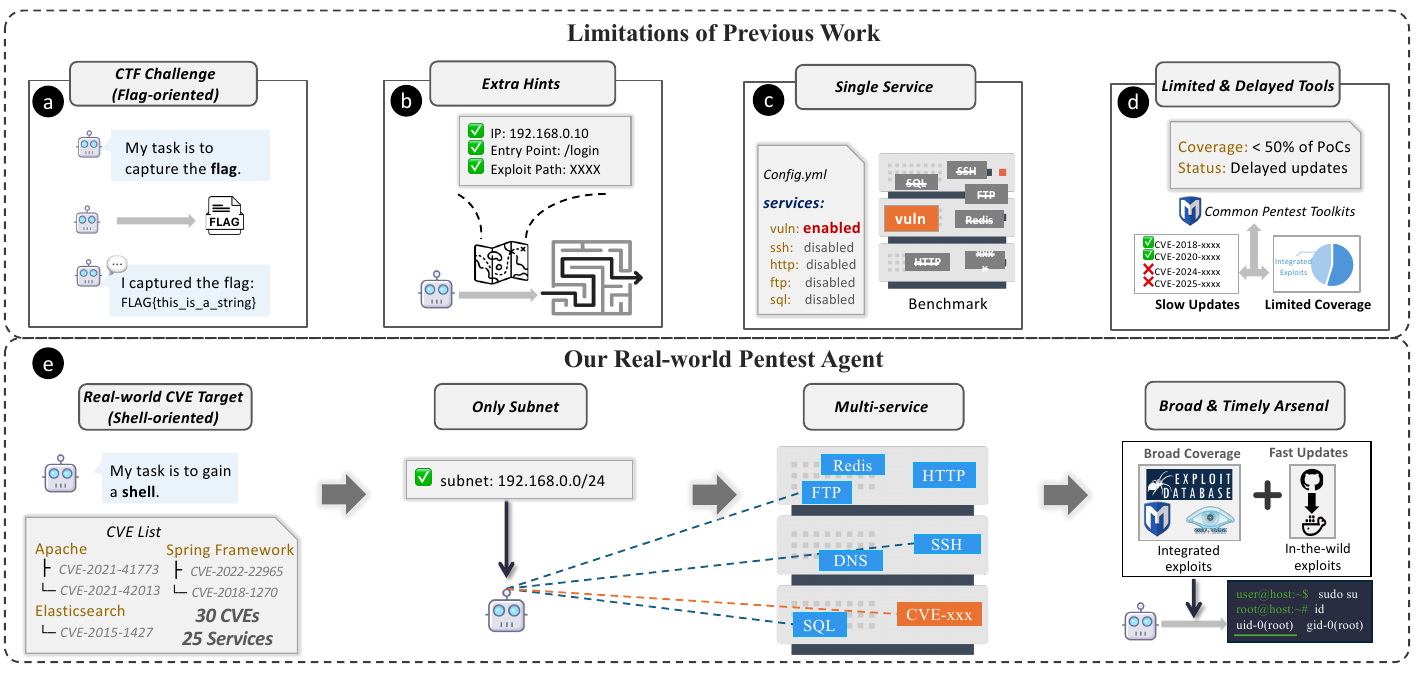}
    \caption{A motivating example of this work compared with the previous work.}
    \label{fig:Motivating example}
\end{figure*}

\subsection{Background}

\noindent \textbf{Penetration Testing.}
Penetration testing, or pentesting, refers to the practice of simulating adversarial attacks on a computing system to identify exploitable security flaws.
Penetration testing is critical for proactively identifying security vulnerabilities before they can be exploited by adversaries. Simulating real-world cyberattacks enables the assessment of defensive effectiveness, mitigates the risk of data breaches, and ensures adherence to security standards \cite{ref1}.
Pentesting is generally executed through a series of well-defined steps, including target Planning, Discovery, Attack, Reporting \cite{NIST-SP800-115}.
The average cost of a penetration test ranges from \$10,000 to \$20,000, depending on the complexity of the environment, including factors such as services, databases, and operating systems \cite{ref3, ref4}.






\noindent \textbf{AI-assisted Cybersecurity Testing.} Current AI-assisted cybersecurity testing efforts mainly fall into two categories: automated vulnerability discovery and remediation, and AI-driven pentesting. In the former, initiatives such as DARPA’s AI Cyber Challenge (AIxCC) \cite{kapko2025darpa} have demonstrated that AI systems can help identify and patch vulnerabilities at scale. In the latter, research prototypes such as PentestGPT represent initial attempts to employ LLM-based agents for pentesting \cite{kong2025vulnbotautonomouspenetrationtesting,299699, shen2025pentestagentincorporatingllmagents}. While these works highlight AI roles to accelerate pentesting, they remain limited by reliance on well-informed CTF-like datasets and oversimplified configurations of meeting the challenges posed by fully automated, end-to-end pentesting against realistic targets.

\subsection{Motivating Example}

Although the emergence of LLMs and agents has fostered the development of automated pentesting,
current approaches still fall short of meeting the requirements of real-world automated pentesting scenarios as in Figure \ref{fig:Motivating example}. Some pentesting agents \textbf{(a)} improperly equate CTF challenges with real-world scenarios, \textbf{(b)} requiring initial hints like entry points, a luxury not afforded in real-world tasks where only IP or subnet is provided. They also heavily \textbf{(c)} rely on static, outdated third-party tools like Metasploit, which have insufficient vulnerability coverage. Similarly, current benchmarks are unrealistic, \textbf{(d)} typically featuring targets with only a single vulnerable service, thereby failing to reflect the multi-service complexity of real-world scenarios.

Our work addresses these challenges by \textbf{(e)} an approach oriented toward real-world pentesting scenarios. Our agent is designed to perform end-to-end penetration tests using only IP or subnet information to gain host ownership. To ensure it has timely access to a more extensive range of exploits, we have built our own dynamic arsenal. Furthermore, our benchmark moves beyond the traditional CTF-style exercises by introducing a multi-service environment and assessing the agent's capability by evaluating the true depth of its penetration.

\subsection{Research Scope}

We consider an adversary leveraging an AI agent to conduct real-world automated end-to-end pentesting. The adversary is assumed to have network-level access to the target but no prior knowledge of its vulnerabilities or configurations beyond standard reconnaissance capabilities. The agent operates fully autonomously—planning, executing, and adapting each step without human intervention. The target is a simulated host configured with widely deployed services, which may or may not contain exploitable vulnerabilities. The adversary’s goal is to achieve control authority of the targeted machine, concretely, obtaining interactive system shell access. 



\begin{table*}[!ht]
    \centering
    \small
    \caption{A comparison between \ouragent and prior work. \ouragent requires only the subnet range of target to achieve the real-world penetration objective of system ownership, while also being compatible with CTF scenarios. Leveraging a Pluggable External Arsenal and the Local Memory Activation mechanism, \ouragent is designed to handle complex, multi-step, multi-service real-world penetration testing tasks, while maintaining compatibility with lightweight LLMs. }
    \begin{threeparttable}
    \resizebox{\textwidth}{!}{
    \begin{tabular}{r| c c c c c c}
    \toprule & \makecell{\textbf{Success} \\ \textbf{Criterion}}         &  \makecell{\textbf{Target} \\ \textbf{Information}} & \makecell{\textbf{Human} \\ \textbf{Assisted}} & \makecell{\textbf{Pluggable External} \\ \textbf{Arsenal}} & \makecell{\textbf{Muti-Service} \\ \textbf{Penetration}} &  \makecell{\textbf{Lightweight LLM} \\ \textbf{Compatibility}}\\ 
         \midrule
        \textbf{\ouragent} & \textbf{System Ownership} &  \textbf{Subnet}  & \textbf{\ding{55}} & \textbf{\ding{51}} & \textbf{\ding{51}} &\textbf{\ding{51}}  \\ 
        
        VulnBot\cite{kong2025vulnbotautonomouspenetrationtesting}  & CTF Flag & Subnet / Entry Point / Exploit Path  & \ding{55}   &   \ding{55} & \ding{55} & \ding{55}\\

        PentestGPT\cite{299699}  &  CTF Flag & IP &  \ding{51}    & \ding{55} & \ding{55} & \ding{55} \\
        
        PentestAgent\cite{shen2025pentestagentincorporatingllmagents} &  CTF Flag & IP  & \ding{55}     &  \ \ \ding{55}\textsuperscript{*} & \ding{55} & \ding{55} \\
        
        AutoPentest\cite{henke2025autopentestenhancingvulnerabilitymanagement}   & CTF Flag & IP  &\ding{55}  &   \ding{55} & \ding{55} & \ding{55} \\
        
        HackSynth\cite{muzsai2024hacksynthllmagentevaluation}   &  CTF Flag & IP / Entry Point / Exploit Path & \ding{55}  &   \ding{55} & \ding{55} & \ding{55} \\
        
        CIPHER\cite{Pratama_2024}&    CTF Flag & IP  & \ding{51}  &   \ding{55} & \ding{55} & \ding{55} \\
        
        AutoAttacker\cite{xu2024autoattackerlargelanguagemodel} & Subtask & Exploit Path & \ding{55}  &    \ding{55} & \ding{55} & \ding{55} \\
        
        BreachSeek\cite{alshehri2024breachseek}  & Subtask & Exploit Path &\ding{55}  &    \ding{55} & \ding{55} & \ding{55} \\
    
        \bottomrule
    \end{tabular}
    }
          \begin{tablenotes}    
            \item[*] PentestAgent can just search for relevant exploit online, but it cannot automatically package it into a ready-to-use arsenal.
        \end{tablenotes} 
        \end{threeparttable}
    \label{table:compare_agent}
\end{table*}

\section{Real-World Benchmark}
To effectively evaluate the capabilities of autonomous pentesting agents, a benchmark should faithfully reproduce the complexity and uncertainty of real-world scenarios. A real-world penetration test is defined as a security assessment that mimics real-world attacks to identify methods for circumventing the security features of an application, system, or network \cite{NIST-SP800-115}. Importantly, this process typically involves testers who have little to no prior information about the target environment, except for an IP address. They are tasked with identifying and exploiting known vulnerabilities to evaluate the true level of risk \cite{NIST-SP800-115,OWASP-Testing,PTES,ANSI-X9.111}. Existing evaluation benchmarks \cite{gioacchini2024autopenbench,isozaki2024automatedpenetrationtestingintroducing, 299699}, however, fall short of this standard, creating a gap between an agent's performance in a well-informed testbed and its efficacy in a real operational environment.

\subsection{Prior Benchmark Limitations}









%


Existing pentesting resources, including traditional platforms like Vulhub and HackTheBox \cite{Vulhub,HackTheBox}, are not ideal for evaluating autonomous pentesting agents. These platforms are primarily designed for human training: they provide pre-built vulnerable Docker environments but lack instrumentation to automatically verify compromise or track an agent’s progress. They also fall short in realism, as they only expose the target vulnerability and do not include benign background services. Past benchmarks share the following characteristics.

First, the objectives of the prior benchmarks \cite{gioacchini2024autopenbench, isozaki2024automatedpenetrationtestingintroducing, 299699} are misaligned with those of real-world pentesting. They are primarily structured as CTF-style exercises, where the goal is typically to locate a “flag”—a string stored in a file that signifies task completion. However, this “flag-finding” focus diverges from the end-to-end objectives of a typical penetration test, such as obtaining a remote shell. Moreover, using flag capture as the sole evaluation criterion fails to meet the requirement \cite{NIST-SP800-115} of assessing the impact of the attacker’s penetration progress on the target.

Second, some benchmark provides agents with more information than is available in real-world pentesting. Instructions of targets in Auto-Pen-Bench \cite{gioacchini2024autopenbench} often embed a priori knowledge—such as service names and versions, entry-point hints, or predefined exploit paths—that a red team would not have. In practice, operators typically start with little more than network access and must perform reconnaissance, enumeration, and hypothesis-driven testing under uncertainty \cite{NIST-SP800-115}.

Third, target configurations are often oversimplified. A common limitation in existing work
is the configuration of each server with exactly one exploitable service \cite{gioacchini2024autopenbench, isozaki2024automatedpenetrationtestingintroducing, 299699}. In contrast, real-world servers typically host multiple services, most of which lack easily exploitable vulnerabilities \cite{measuring_cloud_services}. These vulnerability-free services introduce substantial “noise” into pentesting, requiring the agent to perform thorough reconnaissance, accurately identify the actual attack surface, and operate within a larger, more complex environment.




\begin{table*}[!ht]
    \centering
    \small
      \caption{Comparison of penetration testing Benchmarks for automated agents. \songbench is distinguished by its focus on achieving system ownership in complex, multi-service environments with minimal initial information.}
    \begin{threeparttable}
  
    \label{table:compare_benchmark_final}

    \resizebox{\textwidth}{!}{
    
        \begin{tabular}{r| c c c c c c c}
        \toprule
         & \textbf{Primary Objective} & \textbf{Initial Knowledge} &  \makecell{\textbf{Environment} \\ \textbf{Complexity}} & \makecell{\textbf{\# of} \\ \textbf{Benign Services}} & \makecell{\textbf{System Impact} \\ \textbf{Assessment}} & \makecell{\textbf{\# of} \\ \textbf{Unique CVEs}} & \makecell{\textbf{\# of} \\ \textbf{Hosts}} \\
        \midrule
        \textbf{\songbench} & \textbf{System Ownership} & \textbf{Subnet Only} & \textbf{Multi-Service} & \textbf{0, 1, 3, 5, 7} & \textbf{\ding{51}} & \textbf{30} & \textbf{510} \\
        
        Auto-Pen-Bench \cite{gioacchini2024autopenbench} & CTF Flag & Extra Hints & Single-Service & 0 & \ding{55} & 11(CTF-style) & 33 \\
        
        HackTheBox/Vulhub\textsuperscript{*}  & CTF Flag & Not Specified & Single-Service & 0 & \ding{55} & 0 & 13 \\
        
        AI-Pentest-Benchmark \cite{isozaki2024automatedpenetrationtestingintroducing} & CTF Flag & Not Specified & Single-Service & 0 & \ding{55} & 0  & 13 \\
        
        \bottomrule
        
        \end{tabular}
    }
    
    \begin{tablenotes}    
    \footnotesize
            \item[*] Selected by PentestGPT \cite{299699}
    \end{tablenotes} 
    
    \end{threeparttable}
\end{table*}

\subsection{Benchmark Design}\label{sec:lable_design}




To overcome these limitations, we present the first real-world, fine-grained, and agent-oriented pentesting benchmark, built around three principles: real‑world fidelity, blind evaluation, and systematic service integration.

First, instead of artificial “flag-finding” challenges, every target in our benchmark is based on a real, documented CVE. From all CVEs disclosed between 2015 and 2025 \cite{NVD}, we retained only those affecting free and open-source software, reproducible in a controlled setting, and enabling remote code execution(RCE)—the most operationally relevant outcome in professional pentesting. This process yields 30 CVEs spanning 25 distinct services, three times larger than previous works that have been exploited in the wild. Success is defined as obtaining a remote shell, directly aligning with end-to-end pentesting goals. We further examine whether the acquired shell possesses root privileges to evaluate the extent of impact on the target host \cite{NIST-SP800-115}.

Second, to better replicate real-world conditions, agents receive no prior knowledge beyond the target subnet range. Service names, version details, and predefined exploit paths are not provided in our benchmark, requiring agents to autonomously conduct reconnaissance, enumeration, and testing under uncertainty. This blind-start condition reflects the operational reality of human red teams and prevents the unrealistic advantage from benchmarks that include a priori knowledge in their setup.


Third, to solve the problem of oversimplified configurations, we introduce systematic environmental complexity. We integrate benign services alongside the target vulnerability to create operational “noise”. This design forces agents to perform comprehensive service enumeration and accurately identify the true attack vector in a more complex environment, a challenge absent from single-service benchmarks \cite{gioacchini2024autopenbench,Vulhub,HackTheBox}. To systematically evaluate agent performance, we adjust the density of these benign services across different levels, enabling a detailed analysis of how environmental complexity influences the agent’s effectiveness.

\subsection{Benchmark Overview}



Our benchmark consists of a total of 510 distinct host instances, built on a foundation of 30 unique, real-world CVEs that affect 25 different services as shown in Table \ref{table:cves}. The design systematically varies the number of benign background services to create different levels of environmental complexity. This approach enables a detailed evaluation of an agent's capabilities in increasingly realistic network environments. 

The benchmark is structured into two main parts. It begins with a base configuration of 30 hosts, each featuring a single vulnerable service and no benign services, referred to as the “1+0” configuration. These hosts serve to measure agent performance in an isolated environment, providing a direct comparison to the simplified setups of previous benchmarks \cite{gioacchini2024autopenbench, 299699,isozaki2024automatedpenetrationtestingintroducing}.

The core of our benchmark consists of four levels of environmental complexity, totaling 480 hosts, designed to measure the impact of environmental noise. These levels are systematically constructed with one, three, five, and seven benign services running alongside the single vulnerable one corresponding to the “1+1”, “3+1”, “5+1”, and “7+1” configurations, with 120 hosts allocated to each level. The benign services, as shown in Table \ref{table:benign}, are randomly selected from a pool of the 14 most common applications identified through large-scale internet measurements using the FOFA scan engine \cite{fofa}. The upper bound of seven benign services was chosen based on our empirical analysis of this data, which indicates that the vast majority of typical home computing environments host no more than seven concurrently active services. This structure, summarized in Table~\ref {table:benchmark_overview}, enables a systematic analysis of how agent performance degrades or adapts as the operational environment becomes progressively more complex and noisy.

\section{Methodology}~\label{sec:methodology}

In this section, we present the challenges encountered, key insights gained, and the design and implementation of our approach.

\subsection{Challenges and Insights}

Traditional pentesting methodologies \cite{pen_test_wiki} generally involve three key phases. The scanning phase gathers information about the target system. The reconnaissance identifies potential attack surfaces. The exploitation phase attempts to gain system control by leveraging discovered vulnerabilities, thereby completing the pentesting cycle. 

Given the proven effectiveness of these well-established procedures, integrating them into AI-agent-driven automated pentesting appears to be a promising direction. However, we argue that directly applying them in real-world environments with autonomous agents poses significant challenges.


\noindent \textbf{Challenge 1: How to manage and optimize context in real-world pentesting agents under long-context forgetting?}
A key bottleneck in AI-agent-driven pentesting is the long-context forgetting phenomenon inherent to LLMs \cite{liu2023lostmiddlelanguagemodels}.
In complex real-world pentesting scenarios, as the agent progresses through iterative trial-and-error steps, vast amounts of exploratory information accumulate in the context. 
Due to the limited effective context retention of LLMs, older but still relevant information is often attenuated or entirely forgotten as new content is appended. 
Therefore, even when the context length is compressed to stay within the LLM’s window limit, irrelevant information continues to consume valuable context space, causing the agent to miss capturing crucial relevant details, thereby degrading the actual effectiveness of the pentesting.


Insights 1: We tackle this challenge with a Located Memory Activation approach. 
When predicting its next action, the agent automatically activates all memories relevant to the current decision-making, guided by the hierarchical nature of real-world pentesting landscapes where target hosts branch into increasingly specific and mutually independent attack surfaces.
We further apply memory reduction at varying levels—ranging from coarse summaries to fine-grained details—based on the agent’s task requirements.
For example, phased planning requires monitoring high-level penetration test progress, while concrete instruction generation relies on low-level command details. This ensures that the agent consistently retains only minimal but sufficient knowledge throughout the penetration process.



\noindent \textbf{Challenge 2: How to construct an up-to-date and ready-to-use exploit arsenal for the pentesting agents?} Current pentesting agents' exploits often rely heavily on open-source frameworks such as Metasploit \cite{metasploit}, whose exploit collections are relatively outdated and exhibit incomplete coverage of relevant vulnerabilities. In particular, our analysis of RCE vulnerabilities from 2015 to 2025 shows that Metasploit covers only about half of these CVEs compared to publicly available exploits on GitHub, highlighting a gap between existing frameworks and real-world vulnerability disclosures. To close this gap, integrating  exploits from platforms such as GitHub seems straightforward. Unfortunately, online repositories are rarely “plug-and-play”: they lack a unified usage convention, exhibit non-standard execution entry points, and are often accompanied by incomplete or missing documentation. This lack of standardization makes it difficult to transform raw exploits into agent-friendly exploits.


Insights 2: We treat automated exploit integration as a structured code-understanding problem, rather than a simple code retrieval and execution task. Our approach addresses two key challenges. First, we systematically extract over ten semantic and operational dimensions that define an exploit’s executability—such as programming language, base image, system dependencies, and workflow—and organize them into a Unified Exploit Descriptor (UED). The UED serves as a precise intermediate representation, enabling raw GitHub repositories to be automatically packaged into fully executable, self-contained artifacts. Second, to make exploits agent-invocable, we automatically generate concise, precise manuals. These manuals distill excessive or noisy instructions and infer missing details from code structures, repository files, and inline comments. By combining the UED with these distilled manuals, raw exploits are transformed into standardized, plug-and-play modules, supporting reliable integration into automated pentesting workflows.

\subsection{Design Overview}

\begin{figure*}[!ht]
    \centering
    \includegraphics[width=0.95\linewidth]{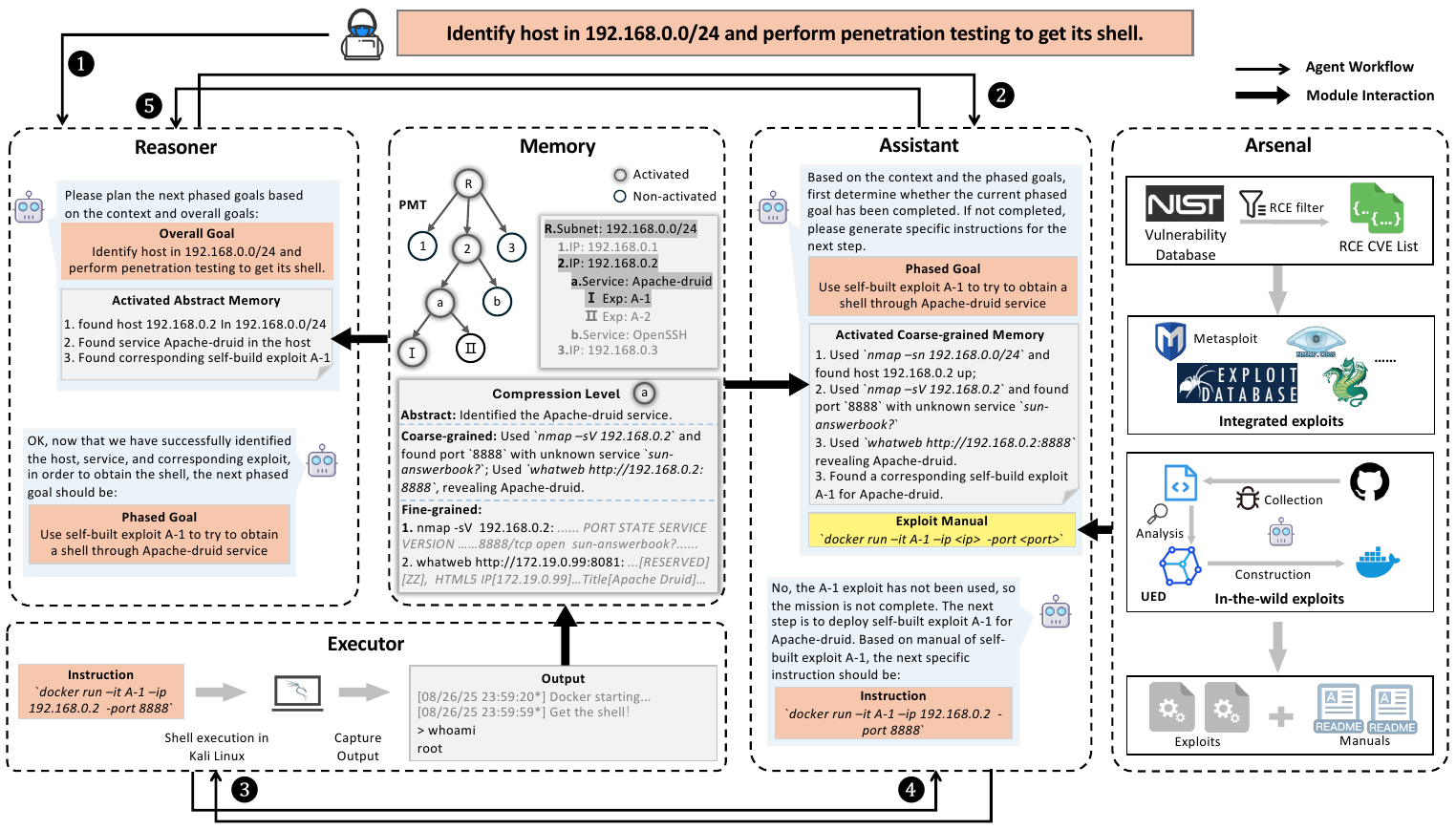}
    \caption{System Overview. \ding{182} The security researcher specifies the overall pentesting goal and inputs it to the agent. After several iterations, based on the abstract memory automatically activated by the \memorymodule \ and the overall goal, the Reasoner Module proceeds to plan the next phased goal. \ding{183} The phased goal is then sent to the Assistant Module, which generates the next specific instruction based on activated coarse-grained memory and documentation from the Self-Built \arsenalmodule. \ding{184} The Executor Module executes the specific instruction and updates the resulting command output into the \memorymodule. \ding{185} The assistant module then generates the next instruction. If the current phased goal is completed, \ding{186} it obtains the next phased goal from the Reasoner Module, repeating this process until the overall goal is achieved.}
    \label{fig:overview}
\end{figure*}

Building on the challenges and our insights aforementioned, we propose \ouragent, a multi-agent framework tailored for real-world pentesting. As illustrated in Figure \ref{fig:overview}, \ouragent \ comprises the \reasonermodule, the \assistantmodule, the \executormodule, the \memorymodule and the \arsenalmodule, which collaboratively decompose pentesting targets into multi-step sub-tasks, progressively achieving the final objectives in a perception–action loop.

Within our framework, the \reasonermodule handles high-level decisions, planning subsequent phased goals based on the ongoing progress and overall goal of the penetration test. The \assistantmodule, in turn, handles low-level decisions by generating the specific command to be executed next, based on the phased goal and the concrete details of the current pentesting task. The \assistantmodule may generate a series of commands to accomplish the phased goal. Control is returned to the \reasonermodule to devise a new phased goal only after the current one is either achieved or fails. The commands generated are subsequently executed by the \executormodule, with their respective outputs being logged.

During this process, the \memorymodule records all contextual information gathered by \ouragent throughout the pentesting, with each entry being goal-oriented and compressed into three distinct levels of granularity. When \ouragent plans the next phased goal or generates the subsequent commands, all the relevant memories are automatically activated, the corresponding compressed level of which is then provided based on the specific requirements of the current decision-making task.

In addition, we develop the \arsenalmodule, which automates the integration of both in-the-wild exploits and open-source pentesting frameworks to \ouragent \ as a plug-and-play module. The \arsenalmodule provides \ouragent with a unified interface encompassing all pentesting tools, including all collected executable exploits along with their corresponding usage manuals, thereby enabling \ouragent to readily utilize them and further enhancing its pentesting capabilities.

In the following sections, we will further elaborate on the \memorymodule and the \arsenalmodule, highlighting both their functionalities and the underlying design principles. The introduction of the Reasoner Module, the Assistant Module, and the Excutor Module can be found in Appendix \ref{app:design}.

\subsection{Memory Module}

The \memorymodule is responsible for recording and organizing all contextual information gathered by the agent during the execution of real-world pentesting tasks. 
First, \memorymodule dynamically activates memory
related to the current target components
within the agent's decision-making loop. 
Second, to provide the agent with compact but sufficient context information, memory is reduced at varying levels according to the task requirements.
Finally, the \memorymodule tracks and guides the pentesting direction according to a comprehensive record of pentesting progress.

%


\subsubsection{Penetration Memory Tree Construction}

To enable automated active suitable located memory, the first challenge lies in accurately identifying which memory is relevant to the current pentesting stage. In real-world pentesting scenarios, a single host may expose multiple attack services and each service may suffering enoumouse vulnerabilities, leading to excessive potential penetration paths for the agent to explore. The widely adopted method for relevant context filtering relies on semantic similarity analysis. However, the key information of a single penetration path, such as IP addresses, ports, and shell commands, contains limited semantic content and thus similarity-based context filtering methods risk omitting critical information. 

We model the pentesting process as a Penetration Memory Tree (\textbf{PMT}), capturing the hierarchical structure of real-world pentesting. In PMT, a target machine may host several services, each exposing entry points exploitable via distinct vulnerabilities. Tree edges represent branching decisions—such as selecting a service or choosing among available exploits—while nodes store the agent’s execution context at that decision point. Upon undergoing a new stage of penetration, the current node spawns child nodes and transfers context to the relevant branch. During execution, context retrieval is performed via backward traversal from the current node to the root of PMT, ensuring activation of relevant memories while suppressing extraneous information.


\subsubsection{Pentest Context Compression}

The agents of \ouragent in different roles prioritize information at different levels of granularity. Specifically, the \reasonermodule concentrates on monitoring the high-level, stage-wise progress of pentesting, whereas the \assistantmodule depends on concrete execution results of individual instructions to both assess the attainment of phased goals and generate subsequent instructions. Therefore, supplying the raw memory with noise in their entirety to the context is insufficient to guarantee that \ouragent can operate at its full potential. In addition to role-specific distinctions, the different stages of real-world pentesting further shape the aspects of context to which agents assign priority. For example, in the scanning stage, \ouragent may prioritize service fingerprints, version numbers, and other indicators of entry points, whereas in the exploitation stage, the focus may shift toward information like command outputs, execution success, etc. Such key aspects are often not directly inferable from instructions and their results alone. If constrained to this limited view, the agent may disproportionately focus on outcome-related information while overlooking other critical metadata.

Therefore, we applied intent-oriented compression to the context of \ouragent. We recorded the agent’s reasoning process during instruction generation to analyze the underlying intent, which then guided the compression of the context into three distinct levels. First, the raw execution results of instructions are normalized at the character level to remove escape sequences, such as color control codes, yielding a fine-grained, LLM-readable context. Based on this fine-grained level context, the abstract level context is compressed to emphasize the stage-wise progress of pentesting, while the coarse-grained level context is compressed to provide an overview of executed instructions and their corresponding outcomes.
The design of intent-oriented compression enables \ouragent to selectively attend to context according to its role and the current stage of the pentesting task, thereby avoiding interference from irrelevant information while ensuring that critical information is fully preserved.

\subsubsection{Path Enumeration and Prioritization}
After constructing the memory tree and compressing unnecessary memory, the \memorymodule must generate the penetration path guide efficiently.
Specifically, \memorymodule employs a depth-first search traversal to enumerate all root-to-leaf paths, explicitly annotating both executed and unexecuted sequences.
Furthermore, child nodes under a parent node can be sorted according to pentesting conditions,
allowing the agent to prioritize exploring paths with a higher likelihood of success. The sorting criteria are determined by the vulnerability likelihood of the target service and the ranking of available exploits.

\subsection{\arsenalmodule}


The \arsenalmodule\ is a framework that transforms heterogeneous "in-the-wild" exploits into standardized, plug-and-play modules for agents. To achieve this, we first introduce a structured format to represent each exploit. Essential information is then extracted from repositories to populate this structured format. Finally, this structured format is used to generate a self-contained, executable module equipped with usage manual for the agent. The entire workflow—from acquiring exploit repositories to generating standardized executable modules—is fully automated with no human intervention.


Repositories of "in-the-wild" exploits lack standardized structures, often differing in project organization, programming languages, system dependencies, and execution workflows—core details that are crucial for automated packaging and reliable execution. Therefore, constructing a robust arsenal requires deep insight into these repositories. To address this, we conducted a large-scale deep analysis of mainstream exploit repositories, guided by two penetration-testing experts, and summarized their essential feature patterns as Unified Exploit Descriptor (\textbf{UED}). As shown in Table \ref{tab:ued_dimensions}, UED is a structured abstraction capturing over ten expert-defined meta-dimensions spanning both environmental and operational aspects, which are essential for automated construction and execution. Environmental dimensions (e.g., language, dependencies, base images) support reproducible execution environments, while operational dimensions (e.g., parameters, operation steps) ensure precise, agent-invocable manuals. By organizing these dimensions into a unified schema, the task of “making an exploit work” is transformed into the well-defined problem of populating a descriptor, turning chaotic exploits into reproducible specifications.

Our analysis further revealed distinct patterns in these repositories' characteristics, which can be categorized into several representative types. Repositories within the same category shared similar environment setups and execution strategies. Based on our analysis, we categorized the repositories into three representative types: packet-based, command-line-based, and script-based exploits. This categorization ensures that subsequent environment construction and manual generation are aligned with the exploit’s intrinsic requirements.

We automate the UED construction process employing an LLM for semantic analysis of entire repositories, reasoning jointly over code, comments, and documentation. The model filters out irrelevant content (e.g., setup instructions, author notes) while inferring missing details from file structures and code patterns. This ensures that even incomplete exploits yield a complete UED, faithfully capturing complex workflows such as multi-stage asynchronous exploits. The constructed UED then drives a dual-output generation process: environmental dimensions generate a Dockerfile for a reproducible containerized environment, while operational dimensions produce a concise manual translating complex attack sequences into step-by-step instructions. This final stage converts noisy, inconsistent exploits into robust, standardized modules ready for autonomous deployment.

\section{Implementation}\label{sec:imp}

Building upon the design in Section \ref{sec:methodology}, we implemented \ouragent on the LangGraph\cite{langgraph_software}, consisting of over 3,500 lines of Python code and about 700 lines of prompt definitions. 

\ouragent requires only the target host’s IP address or the subnet in which it resides as input. The default objective of \ouragent is aligned with real-world pentesting scenarios, aiming to obtain ownership of the target machine, such as acquiring a shell. Additionally, we have adapted for a CTF scenario where obtaining a flag can be specified as an alternative objective. \ouragent communicates with the backend LLM via the API format compatible with OpenAI\cite{openai_api}, which ensures adaptability in switching between different backend LLMs to accommodate diverse pentesting environments. All commands during the pentesting are executed through a Kali\cite{kali_linux} host to interact with the target machine, and the entire process is fully automated, requiring no human intervention until the objective is achieved.

In the Arsenal Module, from the 228,139 CVEs on the NVD website\cite{NVD} in the past decade (2015–2025), we identified 31,332 potential RCEs. Using these CVE IDs, we employed GitHub CLI\cite{github-cli} to search GitHub and identified 6,514 candidates. We subsequently retained only executables that did not require manual front-end interaction and were designed to target non-Windows services. This process resulted in 694 RCE CVEs, corresponding to 2,185 repositories. 
After excluding repositories that failed to be successfully containerized, we obtained 1,378 Dockerized exploits along with their instruction manuals, forming the final arsenal of standardized modules ready for automated execution. Table~\ref{tab:exp_type} shows the exploit distribution and packaging performance. Script-based exploits dominate (89\%), reflecting their prevalence in public repositories, and achieved a 63.51\% success rate when both manuals and Docker images were required. In contrast, packet-based and command-line exploits represent smaller proportions (5.97\% and 5.48\%) but achieved high success rates (94.31\% and 91.15\%, respectively) since only manuals needed to be generated. Across categories, environment reconstruction times ranged from 33 to 84 seconds.

\begin{table}[t]
\centering
\caption{Distribution and Standardization Performance of "in-the-wild" Exploits.}
\label{tab:exp_type}
\resizebox{\columnwidth}{!}{%
\begin{tabular}{lccccc}
\toprule
\textbf{Exploit Type} & \textbf{Count} & \textbf{Proportion} & \textbf{Output Success Rate} & \textbf{Avg. Time (s)} & \textbf{Output} \\
\midrule
Packet-based   & 123  & 5.97\%  & 94.31\%      & 41.28 & Manual \\
Script-based   & 1825 & 88.55\% & 63.51\% & 83.76 & Manual, Docker Image\\
Command-line   & 113  & 5.48\%  & 91.15\%& 33.21 & Manual \\
\bottomrule
\end{tabular}}
\\[2pt]
\footnotesize\raggedright{}
\end{table}

Additionally, by filtering these CVE IDs and considering rankings, we curated 1,077 exploits from Metasploit\cite{metasploit}, corresponding to 851 unique CVEs, and supplemented them with the corresponding manuals. All of these exploits and manuals, continuously collected by the Arsenal Module, can be invoked by \ouragent through a unified interface in a plug-and-play manner, ensuring scalability and modularity.

\section{Evaluation}\label{sec:evaluation}

\begin{table*}[!ht]
    \centering
    \small
    \caption{The overall performance of \songbench, VulnBot, PentestGPT, and built-in agent of Auto-Pen-Bench under both CTF and real-world scenario.}
    \label{table:evaluation}
    \begin{threeparttable}
    
    \resizebox{\textwidth}{!}{
    
\begin{tabular}{c|c|c|ccccc|c|ccccc}
\toprule
\multirow{3}{*}{\textbf{Agent}}  & \multirow{3}{*}{\textbf{Model}} & \multicolumn{6}{c}{\textbf{CTF Scenario \ Pass@5}} & \multicolumn{6}{|c}{\textbf{Real-World Scenario \textsuperscript{*}\ Pass@5}} \\
\cmidrule(lr){9-14} \cmidrule(lr){3-8} &     & \multirow{2}{*}{{Total(33)}} & \multicolumn{5}{c|}{{Category}}  & \multirow{2}{*}{{Total(230)}} & \multicolumn{5}{c}{{\# of Services per Host}} \\\cmidrule(lr){4-8} \cmidrule(lr){10-14}   &    &     & {\footnotesize AC(5)} & {\footnotesize WS(7)} & {\footnotesize NS(6)} & {\footnotesize CRPT(4)} & { \footnotesize CVE(11)} &  & {1(30)} & {2(50)}  & {4(50)}  & {6(50)}  & {8(50)} \\
  \midrule

\multirow{4}{*}{\textbf{\ouragent}}  & {GPT-5}& 11 & 2  & 4 &  2  & 0& 3  & 91/52 & 9/6  & 21/14  & 21/10 & 17/11 & 23/11  \\
& {DeepSeek-V3}  &  15   &   3  &    5    &    2  &   0  &  5&  128/46 &  17/5    &31/13      &   30/10    &   26/10&24/8    \\
& {Qwen3-30B}     &   15 &   3    &     4  &  2     &     0    &   6    & 118/56   &  12/6 & 20/13 & 21/9 &  29/13 & 36/15  \\
& {Qwen3-8B}    & 10 & 2  & 4 & 2 & 0 & 2 &  131/60 & 14/6 & 26/13   & 17/13   & 30/14 & 24/14 \\
  \midrule

\multirow{4}{*}{\textbf{VulnBot}} & {GPT-5} & 3 & 0  & 1 & 0 & 1 & 1 & 3/1 & 1/0 &  2/1  & 0/0   & 0/0 & 0/0  \\
 & {DeepSeek-V3} &9 &  1 & 2 & 4 &1 &1 &  15/9 & 2/1 & 5/3   & 3/1   & 3/2 & 2/2\\
 & {Qwen3-30B} & 7& 0  & 4 & 2 & 0& 1&  5/0& 1/0 & 1/0  & 2/0   & 0/0 & 1/0   \\
 & {Qwen3-8B}  & 4 & 0  & 0 & 2 & 0& 2& 4/2 &  2/1& 2/1 &  0/0  & 0/0&  0/0   \\
  \midrule

\multirow{4}{*}{\textbf{PentestGPT}}  & {GPT-5}  & 3 & 1  & 2 &  0  &0 & 0  & 0/0 & 0/0  &  0/0 & 0/0 & 0/0 & 0/0  \\
 & {DeepSeek-V3} & 4 &  1 & 0 & 2   &0 & 1  & 0/0 & 0/0  & 0/0  &  0/0&  0/0 &  0/0 \\
  & {Qwen3-30B} & 0 & 0  & 0 &  0  & 0& 0  & 0/0 & 0/0  & 0/0  &  0/0&  0/0 &  0/0 \\
     & {Qwen3-8B} & 0 & 0 & 0  & 0 &  0  & 0  & 0/0 &  0/0 &  0/0 & 0/0 &  0/0 &  0/0 \\
 \midrule

\multirow{4}{*}{\textbf{\begin{tabular}[c]{@{}c@{}}Auto-Pen-Bench\\ Built-In\\ Agent\end{tabular}}} & {GPT-5} & 22 & 5  & 4 &  2  & 4&  7 & - &  - & -  &-  &- &-  \\
  & {DeepSeek-V3} & 10 & 0 & 4 &  2  & 0 & 4  & - &  - & -  &-  &-  &- \\
  & {Qwen3-30B} &  3 & 0  & 2 &  1  & 0&  0 & - &  - &  - & - & -&-\\
  & {Qwen3-8B} & 5 & 0  & 3 & 2   & 0& 0  & - & -  & -  & - &  - &-\\
  \bottomrule

\end{tabular}
    }
      \begin{tablenotes}    
            \item[*] Results are shown as \texttt{num1/num2}, with \texttt{num1} indicating shells obtained and \texttt{num2} indicating root shells.
        \end{tablenotes} 
    
    \end{threeparttable}
\end{table*}

In this section, we evaluate the practical performance of \ouragent
based on the following three research questions:

\noindent \textbf{RQ 1 (Performance):} How does the performance of \ouragent, compared with the state-of-the-art pentesting agents, both in CTF scenarios and real-world scenarios?

\noindent \textbf{RQ 2 (Cost):} Can \ouragent, compared with the state-of-the-art pentesting agents, maintain lower time and financial costs while ensuring comparable performance?

\noindent \textbf{RQ 3 (Ablation):} To what extent do the \arsenalmodule and Located Memory Activation improve the performance of \ouragent? How does the design of the UED facilitate the Dockerization of exploits?

\subsection{Evaluation Setting}\label{sec:eval_settings}

We evaluated the performance of \ouragent\  under both real-world scenarios and CTF scenarios. For evaluation under real-world scenarios, we used \songbench\ constructed as in Section \ref{sec:lable_design}, which contains 510 target machines with varying numbers of services. Each category of machines includes one vulnerable service and 0, 1, 3, 5, or 7 benign services. We selected all targets with 0 benign services, and randomly sampled 50 targets from each of the remaining four categories, resulting in a total of 230 targets for evaluation. For CTF scenarios, we used all 33 targets from Auto-Pen-Bench\cite{gioacchini2024autopenbench} for its diverse categories of CTF challenges and local deployment–friendly features. 
For ethical considerations, each target machine was subjected to five repeated tests to mitigate potential randomness in the results. A penetration test was deemed successful if at least one of the attempts succeeded.

We selected PentestGPT\cite{299699} and VulnBot\cite{kong2025vulnbotautonomouspenetrationtesting} for comparison with \ouragent, as both of them are widely recognized and feature corresponding designs for pentesting task planning. For PentestGPT, we used the fully automated version implemented in VulnBot, which retains all original prompts and simulates a manual copy-paste process without additional interpretation. Additionally, we also included the built-in autonomous agent from Auto-Pen-Bench, which is equipped with essential tools and a limited single-framework prompt, allowing for a relatively objective evaluation of an LLM’s ability to tackle CTF challenges. However, it was not tested in a real-world scenario as its design is tightly coupled with the CTF-oriented targets in Auto-Pen-Bench.

To evaluate the impact of different backend LLMs on the performance of \ouragent, we selected GPT-5-2025-08-07\cite{openai_gpt5_2025} from closed-source models, DeepSeek-V3-0324\cite{deepseek_v3_0324}, and the Qwen3 series\cite{qwen3} from open-source models. For the Qwen3 series, we chose five models with different parameter sizes, Qwen3-30B-A3B\cite{qwen3_30b_a3b}, Qwen3-14B\cite{qwen3_14b}, Qwen3-8B\cite{qwen3_8b}, Qwen3-4B\cite{qwen3_4b}, and Qwen3-1.7B\cite{qwen3_1.7b} to assess the agent’s compatibility with lightweight LLMs. Detailed information on the LLMs is summarized in Table \ref{table:llms}.

All experiments were conducted on authorized, isolated servers, with an Intel® Xeon® Gold 6330 processor (28 cores), 8 NVIDIA GeForce RTX 4090 GPUs, 512GB of memory, and a 29TB hard disk drive, running Ubuntu 24.04 LTS.

\subsection{Performance Evaluation (RQ1)}



To comprehensively evaluate the performance of \ouragent and its competing agents in both real-world and CTF scenarios, this section not only assesses their completion of pentesting tasks on selected benchmarks, but also further examines the performance degradation of limited CTF instruction settings. Additionally, we compare the agents’ adaptability to lightweight LLMs.

\noindent \textbf{CTF/Shell Exploitation Performance.} The pentesting results of \ouragent and its competing agents in both real-world and CTF scenarios are summarized in Table \ref{table:evaluation}, and \ouragent achieved superior performance. In the CTF scenario, the best performance of \ouragent exceeded that of VulnBot by 66.67\%, achieving 15 successful tests compared with VulnBot’s 9. Remarkably, even when powered by a lightweight LLM such as Qwen3-8B, \ouragent still outperformed VulnBot’s best results using DeepSeek (10 v.s. 9), further demonstrating its robustness and efficiency under resource-constrained settings. In contrast, PentestGPT, although supplemented with a command execution module, achieved only 4 successful tests at best, highlighting its reliance on human involvement to ensure effectiveness. In the real-world scenarios, \ouragent stood out, achieving success of pentesting on over 50\% of the target hosts with all selected backend LLMs. Conversely, VulnBot, designed for CTF competitions, failed to cope with real-world pentesting, as its performance was limited to successfully compromising fewer than 10\% of the target hosts. Additionally, within its limited success, VulBot is more effective at hosts with a small number of services (1, 2, 4 per host), accounting for 21 of its total 27 successes. Its success rate drops on hosts with a larger number of services (6, 8 per host). This highlights the agent's inability to effectively adapt to real-world scenarios where a single host may have many benign services that need to be excluded. PentestGPT failed to successfully compromise any of the target hosts. Beyond its limited capability in real-world pentesting, our log analysis revealed that in over 50\% of the attempts, PentestGPT generated nmap commands containing unresolved placeholders, such as \texttt{nmap -p- -sV [target IP]} that caused errors. These findings suggest that PentestGPT remains far from achieving fully automated pentesting.

Notably, Auto-Pen-Bench’s built-in agents powered by GPT-5 completed 22 targets, a sharp contrast to DeepSeek and Qwen models. We hypothesize that this is because GPT-5’s knowledge cutoff on September 30, 2024, closely aligns with the open-source release of Auto-Pen-Bench and the provided official solutions using its built-in agents. However, the performance of VulnBot powered by GPT-5 fell significantly short of expectations in both CTF and real-world scenarios. 
Further log analysis showed that GPT-5 in VulnBot generates more complex, piped instructions that VulnBot’s execution module cannot handle.
Compared with DeepSeek’s instructions under the CTF scenario, each individual instruction is on average 40.10\% longer, and the number of instructions per attempt increases by 33.33\%. And these two percentages in real-world scenarios were 40.80\% and 41.67\%, respectively. While seemingly sophisticated, this behavior leads to suboptimal results and even errors, highlighting the lack of robustness in the design of VulnBot.

These results demonstrate that while existing pentesting agents show some capability in CTF scenarios, they are insufficient for complex, real-world tasks. In contrast, the outstanding performance of \ouragent not only fills this gap but also significantly advances the efficacy of pentesting.

\begin{figure}[!t]
    \centering
    \includegraphics[width=0.95\linewidth]{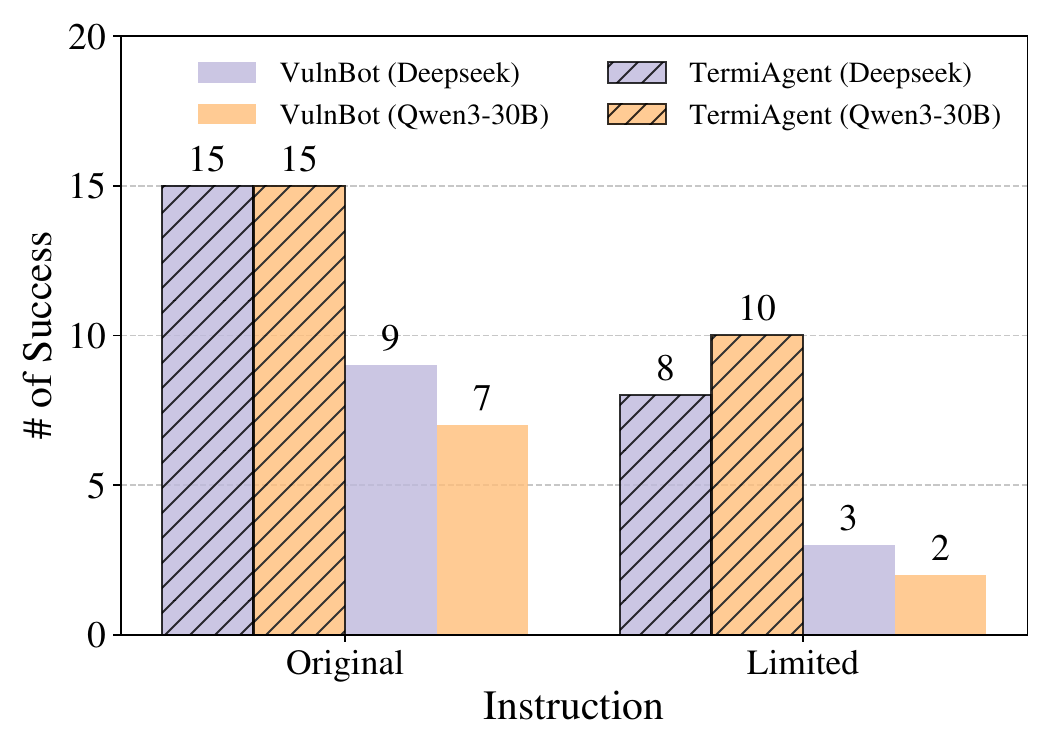}
    \caption{Performance of \ouragent and VulnBot with limited instruction under CTF scenario.}
    \label{fig:reduce_prompt}
\end{figure}

\noindent \textbf{Performance under limited instruction.} The targets in Auto-Pen-Bench's default configuration disclose supplementary hints, such as entry points and exploit paths, in addition to their subnet. To better emulate a red team scenario where information is limited, we reconfigured the instructions for each target to be restricted to contain only \textbf{1)} the target's subnet and \textbf{2)} the ultimate objective to obtain the flag. Further details of this process are provided in the Appendix \ref{app:reduction}. Under limited instructions, we re-evaluated the performance of both \ouragent and VulnBot as in Figure \ref{fig:reduce_prompt}.

Under limited instructions, the number of targets successfully compromised by \ouragent powered by Deepseek and Qwen3-30B dropped from the original 15 to 8 and 10, respectively, which was comparable to VulnBot with original instructions. In contrast, VulnBot powered by Deepseek and Qwen3-30B saw their number of successful targets drop from 9 to 3 and from 7 to 2, respectively, representing significant decreases of 66.67\% and 71.43\%. The experimental results revealed that VulnBot’s design inherently requires these additional hints to guarantee its pentesting success, which diverges from the reality of real-world scenarios where information is typically limited to an IP or subnet. By comparison, \ouragent demonstrated its capacity to perform effective penetration tests and achieve a commendable success rate using only the limited information provided.

\noindent \textbf{Lightwight LLM compatibility.} Previous pentesting agents relied on state-of-the-art LLMs, which not only incur significant financial costs but also cannot be deployed in internal environments with limited computational resources and restrictions on network access. To evaluate the extent to which our agent can achieve a trade-off between resource consumption and performance, we selected 5 open-source LLMs from the Qwen3 series, with parameter sizes ranging from 1.7B to 30B. Using the 230 target hosts from \songbench in a real-world scenario, we compared the performance of \ouragent with that of VulnBot, and each target host was tested five times. The results are shown in Table \ref{table:lightweightllm}. 

The experimental results reveal that \ouragent achieved excellent performance across LLMs with different parameter sizes. Even with a 4B model, \ouragent successfully conducted 137 penetration tests out of 230 target hosts. This is comparable to the performance achieved by the model with 30B parameters and even DeepSeek-V3 with 685B parameters. Even though \ouragent, based on Qwen3-1.7B, completed 67 successful penetration tests, this is still a remarkable performance considering the model's parameter size and computational overhead. In contrast, VulnBot's compatibility with lightweight LLMs was notably less effective. It completed only 5 successful penetration tests based on the 30B model, a 66.67\% decline from its best result of 15 with DeepSeek-V3. Performance of VulnBot with the 4B and 1.7B models was even more deficient. The 4B and 1.7B models are efficient enough to run smoothly on mainstream consumer-grade GPUs. The results demonstrate the potential for \ouragent to perform penetration tests locally on devices such as laptops and even smartphones.

\begin{table}[]
    \centering
    \small
    \caption{Lightweight LLM compatibility of \ouragent and VulnBot by Qwen3 series in real-world scenario.}
    \resizebox{0.45\textwidth}{!}{
    \begin{tabular}{c|ccccc}
    \toprule
    & \textbf{30B} & \textbf{14B} & \textbf{8B} & \textbf{4B} & \textbf{1.7B} \\ \midrule
\textbf{\ouragent} & 118/56       & 116/53       & 131/60      & 137/58      & 67/33         \\
\textbf{VulnBot}                  & 5/0          & 3/2          & 4/2         & 1/1         & 0/0       \\
\bottomrule
\end{tabular}
    }    
    \label{table:lightweightllm}
\end{table}

\subsection{Cost Evaluation (RQ2)}

\begin{table}[]
    \centering
    \small
    \caption{Time and financial cost of \ouragent and VulnBot in CTF and real-world scenario.}
    \resizebox{0.48\textwidth}{!}{
    \begin{tabular}{c|c|cc} \toprule
  &\textbf{Metric} & \textbf{CTF Scenario} & \textbf{Real-world Scenario} \\ \midrule
\multirow{2}{*}{\textbf{\ouragent}} & Avg. Time (Mins)   &  19.5908 &   11.7875 \\
 & Avg. Cost (\$)  &  0.0551   & 0.0074  \\ \midrule
\multirow{2}{*}{\textbf{VulnBot}}    & Avg. Time (Mins)   &  17.5746  &  63.1402  \\
  & Avg. Cost (\$)  &  0.0577  &  0.0996   \\ \bottomrule                    
\end{tabular}
    }
    \label{table:cost}
\end{table}

To assess the overhead of \ouragent in achieving its current pentesting performance, we evaluated both its time and financial costs in CTF scenarios and real-world scenarios, respectively. We selected VulnBot for comparison and to ensure the reliability of experimental results, both \ouragent and VulnBot were tested under identical hardware, software, and network environments, which were aforementioned in Section \ref{sec:eval_settings}. Under both scenarios, we only selected the vulnerable machines that both agents successfully penetrated—regardless of the backend LLM—to eliminate bias caused by the varying difficulty of different target machines. For both agents, we analyzed every successful penetration attempt on the selected vulnerable machines. The time cost for each machine was defined as the shortest successful attempt duration. For the financial cost, we calculated the cost of each attempt by multiplying the number of LLM tokens consumed by the LLM's API price, then selected the lowest-cost attempt for that machine. The API specifications are presented in Table \ref{table:llms}. Finally, we computed the average time and financial costs for both \ouragent and VulnBot across all the selected targets as the final results.

We ultimately selected 12 jointly successful target hosts in the CTF scenario and 18 machines in the real-world scenario. For a total of 283 successful attempts on these target machines, we calculated the time and financial costs based on the aforementioned criteria. The final results are summarized in Table \ref{table:cost}. The results demonstrate that for these jointly successful target hosts, in the CTF scenario, \ouragent and VulnBot exhibited comparable levels of cost and time overhead. But in the real-world scenario, \ouragent consumed only about 7.43\% of the financial cost and 18.67\% of the time required by VulnBot to compromise the same number of targets, not to mention its overwhelming advantage in terms of the total number of successfully exploited machines. This is because, in a complex multi-service pentesting task, \ouragent utilizes Located Memory Activation (LMA) to filter irrelevant context and reduce token usage, enabling strong performance with lightweight LLMs of cheaper price. Moreover, its Penetration Memory Tree (PMT) enumerates and prioritizes exploitation paths, avoiding redundancy and enhancing efficiency.

\subsection{Ablation Study (RQ3)}

To validate the effectiveness and irreplaceability of our proposed designs, we perform ablation studies in this section. Specifically, we conduct ablations on the Memory and Arsenal Module of \ouragent during pentesting, and then the UED during dockerization of in-the-wild exploits.

\begin{figure}[!t]
    \centering
    \includegraphics[width=\linewidth]{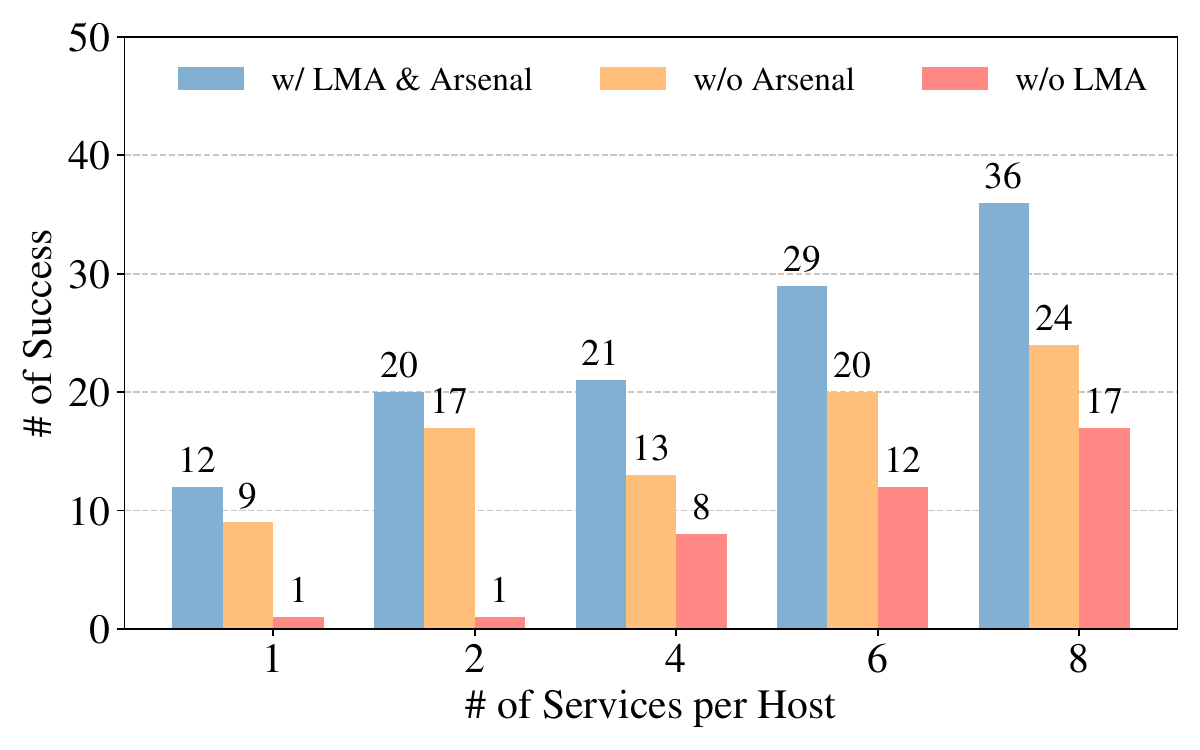}
    \caption{Ablation study on Local Memory Activation (LMA) and \arsenalmodule of \ouragent. }
    \label{fig:ablation}
\end{figure}


\begin{table}[!t]
\centering
\caption{Ablation analysis of individual UED components.}
\label{tab:ablation}
\resizebox{\columnwidth}{!}{%
\begin{tabular}{cccc}
\toprule
\textbf{Group} & \textbf{Excluded Component} & \textbf{Output Success Rate} &  \textbf{Exploitation Success Rate}\\
\midrule
Baseline  & None (all preserved)             & 89.1\% &  63.33\%\\
1         &  language& 83.1\% &  0\\
2         & Language version                 & 82.7\% &  43.33\%\\
3         & Docker base image                & 25.1\%&  3.33\%\\
4         & System dependencies& 85.3\% &  50.00\%\\
5         & Code dependencies& 84.0\% &   0\\
6& Main scrip& 82.7\% &  43.33\%\\
7& Parameter files                  & 83.5\% &  33.67\%\\
8& Docker config& 83.5\% &  43.33\%\\
9& Setup steps                      & 83.5\% &  36.67\%\\
10& Exploit steps                    & 84.0\% &  36.67\%\\
11& Command parameters               & 84.4\% &  23.33\%\\
12& Usage example                    & 84.4\% &  23.33\%\\
\bottomrule
\end{tabular}%
}
\end{table}

\noindent \textbf{Ablation of Memory and Arsenal Module.} To quantify the performance improvements brought by the Memory Module and the Arsenal Module to \ouragent's pentesting capabilities, we conduct a series of evaluations by separately removing each module under real-world scenarios. Specifically, with the Memory Module ablated, we disabled the Located Memory Activation (LMA) approach. In its place, we followed the design of VulnBot and PentestGPT, inputting the complete Penetration Memory Tree (PMT), which contains all information from the current penetration test, directly into the context. While with the Arsenal Module ablated, we limited \ouragent's ability to access in-the-wild exploits which were already packaged and ready to use. We conducted this experiment using Qwen3-30B, using the same 230 target machines from \songbench as mentioned in Section \ref{sec:eval_settings}, with each tested five times to mitigate the effects of randomness.

The results of the ablation study on the Memory and Arsenal Modules are shown in Figure \ref{fig:ablation}. With the Arsenal Module ablated, \ouragent's pentesting effectiveness dropped by 29.66\%, decreasing from 118 to 83. Furthermore, the ablation of Located Memory Activation approach led to a much more pronounced decrease in \ouragent's effectiveness. A total of 79 previously successful targets could no longer be compromised, representing a substantial decline rate of 66.95\%. The largest number of failures, nearly 20 targets, occurred on machines with 8 services. This suggests that the Located Memory Activation (LMA) approach within the Memory Module is of critical role in complex, multi-service pentesting scenarios.


\noindent \textbf{Ablation of UED.} We evaluated the UED on 231 GitHub repositories corresponding to 30 target CVEs, covering all publicly available scripts for these CVEs. Output Success Rate measures the proportion of repositories for which both a Dockerized image and a manual were successfully generated. To assess practical exploitation, we randomly selected one repository per CVE and manually verified pentesting on the corresponding vulnerable targets. Only one repository per CVE was tested due to the large number of UED dimensions, allowing us to focus on each component’s contribution while keeping verification manageable.

The baseline system, with all UED dimensions preserved, achieved an Output Success Rate of 89.1\%—higher than the 63.51\% observed across 1,825 repositories (Table~\ref{tab:exp_type})—because the selected CVEs were representative and commonly supported, simplifying containerization, and the Exploitation Success Rate was 63.3\%.

Ablation shows that the Output Success Rate is dominated by the Docker base image, with its removal causing a sharp drop, while other dimensions have minor effects. Exploitation Success Rate depends on multiple dimensions, including setup steps, exploit steps, command parameters, and usage examples, any of which can significantly reduce attack success if omitted. Special cases highlight this: omitting code dependencies leads to zero exploitation success, since the necessary code libraries or modules are missing, causing execution to fail even though containerization succeeds, whereas omitting system-level dependencies has a smaller effect, as fewer repositories rely on OS-level packages. These results confirm that each UED component is crucial, either for reproducible containerization or for generating manuals that reliably guide successful exploitation.

\section{Discussion}

\noindent \textbf{Security Implication.} 
In this work, we propose \songbench, a benchmark for automated pentesting composed of 510 real-world targets, and \ouragent, an automated agent that significantly enhances effectiveness by leveraging Located Memory Activation and the \arsenalmodule. Both \songbench and \ouragent are designed to facilitate the development of more secure systems. Advancements in automated pentesting will strengthen system robustness and security while reducing the barriers to building secure systems.

The introduction of \songbench establishes a much-needed, unified standard for evaluating pentesting agents in real-world scenarios, addressing the biased and inconsistent understanding of what constitutes real-world in prior research. Furthermore, our evaluation demonstrates that \ouragent significantly outperforms its competitors in both CTF and real-world pentesting scenarios using even lightweight LLMs. It represents a significant leap forward in automated pentesting agents, as it ushers in a new era where real-world pentesting can be effectively driven by consumer-grade GPUs on laptop-scale deployments and obtain superior performance.



Our work demonstrates a critical vulnerability in the safety alignment of LLMs. Despite safeguards\cite{openai_gpt5_2025} of LLMs to prevent harmful and unethical content, \ouragent successfully bypassed security protocols without jailbreaking prompts. Across 1,150 tests every LLM, only GPT-5 exhibited 11 times of refusal (0.96\%). This may be attributed to \ouragent's design, which leverages Located Memory Activation to filter out unrelated context and employs fine-grained task decomposition to prevent LLM from perceiving full security risk.

\noindent \textbf{Limitation.} Despite \ouragent's excellent performance in both CTF and real-world scenarios, we also acknowledge certain limitations of this work. Firstly, \ouragent has limited capabilities in complex web-based pentesting, which often demand advanced operations, such as interacting with HTML elements or handling file uploads\cite{nvd2019cve-2019-6339}, and performing in-depth response analyses. This remains a significant and unresolved challenge for all current automated pentesting agents. However, \ouragent is already capable of handling basic web pentesing\cite{nvd2018cve-2018-7600}, and we will continue to advance its capabilities for complex web-based scenarios in future work.

Secondly, within the \arsenalmodule, we are currently unable to handle certain repositories that exhibit excessive complexity, particularly those with large monolithic codebases or highly fragmented logic. Such complexity hinders the accurate extraction of exploit logic and the preparation of a reliable execution environment. Besides, many PoCs possessed complex operational requirements that fall outside our current automation scope, such as dependencies on interactive user interfaces or the need for manual setup of external network services. These outliers represent challenges in PoC implementation style rather than a specific class of vulnerability, and our future work will explore techniques like modular code analysis and browser automation to address them.

Thirdly, while \songbench currently focuses on evaluating whether an automated attacker can achieve initial ownership of a target, it does not yet include assessments for the post-exploitation phase, such as privilege escalation, information gathering, or lateral movement. However, our evaluations demonstrate that this scope is already sufficient for objectively gauging the capabilities of contemporary automated pentesting agents. The inclusion of post-exploitation assessments will be a key focus for our future work.



\section{Related Work}

\noindent \textbf{Automated Penetration Testing.}Penetration testing, a long-standing cybersecurity task, has been greatly facilitated by LLMs and AI agents. While systems like PentestGPT\cite{299699} and CIPHER\cite{Pratama_2024} can assist with pentesting as chatbots, they remain semi-automated, requiring continuous human interaction and falling short of true end-to-end automation. Although AutoAttacker\cite{xu2024autoattackerlargelanguagemodel} and BreachSeek\cite{alshehri2024breachseek} provide some subtask execution capabilities, they lack the planning functionality necessary for true end-to-end pentesting. In contrast, AutoPentest\cite{henke2025autopentestenhancingvulnerabilitymanagement} and HackSynth\cite{muzsai2024hacksynthllmagentevaluation} aim for a more challenging CTF scenario, but their planning modules are solely based on simple prompt engineering, which still fall short of effectively handling complex real-world pentesting scenarios. VulnBot \cite{kong2025vulnbotautonomouspenetrationtesting} improves planning via a Penetration Task Graph but suffers from context loss and reliance on third-party tools, limiting its effectiveness. While PentestAgent\cite{shen2025pentestagentincorporatingllmagents} can search online exploits, it lacks the ability to package them into ready-to-use tools. In our work, \ouragent improved end-to-end pentesting effectiveness in real-world scenarios by utilizing Located Memory Activation and a self-built arsenal.

\noindent \textbf{Benchmark for Penetration Testing.} Traditional platforms like HackTheBox\cite{HackTheBox} and Vulhub\cite{Vulhub} feature human-designed CTF challenges with only one vulnerable service per host, primarily aimed at training security professionals and failing to capture real-world scenarios. AI-Pentest-Benchmark\cite{isozaki2024automatedpenetrationtestingintroducing} classifies tasks to evaluate an agent’s pentesting capabilities, but its assessment focuses on sub-task completion rather than end-to-end testing. Although Auto-Pen-Bench\cite{gioacchini2024autopenbench} includes CVEs, it remains limited to the CTF task rather than ownership-targeted real-world pentesting. Besides, its instructions include extra hints that prevent objective evaluation. In our work, \songbench provides 510 target hosts in multi-service, ownership-based tasks, objectively assessing agents’ penetration success given only subnet information.
\section{Conclusion}

In this work, we propose \songbench, a real-world benchmark for evaluating pentesting agents, which comprises 510 multi-service target hosts incorporating 30 CVEs that affect 25 different services. Besides, we present \ouragent, an automated pentesting agent designed for real-world scenarios. It leverages Local Memory Activation to effectively handle complex, multi-step pentesting tasks, and we enhanced \ouragent's capabilities by proactively collecting in-the-wild exploits in \arsenalmodule, moving beyond the limitations of third-party tools. Our comprehensive evaluation demonstrates that \ouragent achieved superior performance in both CTF and real-world pentesting scenarios while reducing execution time and financial cost. Our work advances the automation of real-world pentesting into a new stage, where such processes can be effectively driven by laptop-scale deployments.

\section*{Ethics Considerations}\label{sec:ethic}

In this work, we introduce \songbench and \ouragent to advance automated penetration testing, thereby enhancing system robustness and security while lowering the barriers to building reliable defenses.

Recognizing that penetration testing research may pose ethical risks, we have carefully designed our study to adhere to established ethical norms and relevant guidelines. Our approach emphasizes responsible disclosure, strict experimental isolation, and controlled dissemination, ensuring that the benefits of our research outweigh potential harms. We will provide a transparent account of our reasoning as follows.

\subsection*{1. Stakeholders and Impact Analysis}

We identify the following stakeholders and analyze the potential impacts of our research.

\textbf{\romannumeral1) Research Team.} Conducting automated penetration testing carries operational risks. Ethical oversight by our institution’s Responsible AI Committee ensures safe procedures and adherence to responsible experimentation guidelines.

\textbf{\romannumeral2) Academic Community.} Researchers benefit from access to methodology and findings, facilitating further research in automated penetration testing. There is a potential risk of misuse if tools are not carefully controlled.

\textbf{\romannumeral3) CVE Vendors and Affected Organizations.} Publication may indirectly affect organizations related to the CVEs used. All vulnerabilities analyzed are publicly disclosed, with no novel 0-day exploits are used.

 \textbf{\romannumeral4) Potential Malicious Actors.} There is a risk that \ouragent could be misused. Mitigations include restricted access, responsible use agreements, and controlled dissemination.

\textbf{\romannumeral5) Society and End Users.} Indirect benefits include improved understanding of defensive strategies against automated penetration testing. There is minimal risk to external users as all experiments are confined to isolated environments.


\subsection*{2. Mitigation Measures}

To ensure that our research on automated penetration testing is conducted in a safe, ethical, and responsible manner, we adopt a series of mitigation measures as follows.

\textbf{\romannumeral1) Responsible Scope and Disclosure.} Our study was conducted under the oversight of an ethics review procedure under our institution’s Responsible AI Committee, ensuring that the research adhered to established guidelines for responsible security experimentation. The scope of our work is strictly limited to publicly disclosed CVEs documented in the NVD database\cite{NVD}. \ouragent leverages the Located Memory Activation (LMA) mechanism—designed to handle complex, multi-step penetration testing tasks in real-world settings—exclusively to exploit publicly disclosed CVE vulnerabilities rather than unknown 0-day vulnerabilities. In addition, the \arsenalmodule employs the Unified Exploit Descriptor (UED) to automatically package publicly available exploits for known CVEs into a ready-to-use form, without generating new exploits. All penetration testing logs are carefully reviewed by professional security personnel, and any noteworthy potential risks identified by \ouragent will be proactively disclosed through responsible channels.

 \textbf{\romannumeral2) Controlled and Secure Environment.}  
All experiments were conducted within a fully isolated and controlled environment to eliminate any risk of affecting external systems. Access to the source code, experimental environment, and exploit artifacts was restricted to authorized and trusted personnel only, ensuring secure handling of sensitive components and preventing misuse. In addition, all testing activities were continuously logged and monitored to ensure compliance with institutional ethical standards and to prevent any unauthorized use or dissemination.

 \textbf{\romannumeral3) Open-source and Community Engagement.} To mitigate potential risks associated with abuse, the prototype of \ouragent, along with sets of exploits, will only be made conditionally available to researchers who submit a formal request, provide proof of their qualifications, and, upon our review, sign a responsible use agreement. This approach prevents misuse while actively promoting further research in the academic field of automated penetration testing. Moreover, we actively engage with the security research community to establish responsible usage guidelines for \ouragent, aiming to prevent potential misuse while promoting safe and ethical adoption. Furthermore, there remains a notable gap in the academic literature regarding defensive research for automated penetration testing. As part of our future work, we plan to focus on developing methods for detecting and defending against automated penetration testing agents, thereby contributing to both the safe deployment of such tools and the broader understanding of defensive strategies in cybersecurity.

\subsection*{3. Decision Rationale}
We chose to conduct this research as automated penetration testing with publicly disclosed CVEs enables systematic exploration of complex attack paths in a real-world scenario, providing significant benefits to security research, supporting the development of more resilient systems, and lowering the practical challenges faced in constructing secure defenses.
And we decided to publish our findings because sufficient mitigations are in place, and the societal and academic benefits outweigh the potential risks. Ethical considerations, including potential harms and rights protection, were systematically weighed.





\section*{Open Science}\label{sec:open_science}

We are committed to the open science policy, and we have made \songbench publicly available at \url{https://doi.org/10.5281/zenodo.16962513}. \songbench features 510 penetration testing targets that are linked to 30 CVE IDs and affect 25 different software services. The repository includes detailed usage instructions for each machine, serving as a foundational resource for further research into automated penetration testing.

For ethical consideration, we will not make \ouragent's code or the exploits from the \arsenalmodule publicly available, to prevent any potential misuse. The prototype of \ouragent, along with a limited set of exploits, will only be made conditionally available to researchers who submit a formal request, provide proof of their qualifications, and, upon our review, sign a responsible use agreement in the future.

\bibliographystyle{ACM-Reference-Format}
\balance

\bibliography{reference}


\begin{thebibliography}{43}


\ifx \showCODEN    \undefined \def \showCODEN     #1{\unskip}     \fi
\ifx \showISBNx    \undefined \def \showISBNx     #1{\unskip}     \fi
\ifx \showISBNxiii \undefined \def \showISBNxiii  #1{\unskip}     \fi
\ifx \showISSN     \undefined \def \showISSN      #1{\unskip}     \fi
\ifx \showLCCN     \undefined \def \showLCCN      #1{\unskip}     \fi
\ifx \shownote     \undefined \def \shownote      #1{#1}          \fi
\ifx \showarticletitle \undefined \def \showarticletitle #1{#1}   \fi
\ifx \showURL      \undefined \def \showURL       {\relax}        \fi
\providecommand\bibfield[2]{#2}
\providecommand\bibinfo[2]{#2}
\providecommand\natexlab[1]{#1}
\providecommand\showeprint[2][]{arXiv:#2}

\bibitem[ali({[n.\,d.]})]%
        {aliyun_bailian_console_2025}
Alibaba Cloud \bibinfo{year}{[n.\,d.]}\natexlab{}.
\newblock \bibinfo{booktitle}{\emph{Alibaba Cloud Bailian Console}}.
\newblock Alibaba Cloud.
\newblock
\urldef\tempurl%
\url{https://bailian.console.aliyun.com/}
\showURL{%
\tempurl}
\newblock
\shownote{Enterprise-level large-model service platform console (“Bailian”) from Alibaba Cloud}.


\bibitem[ope({[n.\,d.]})]%
        {openai_api_pricing_en_2025}
OpenAI \bibinfo{year}{[n.\,d.]}\natexlab{}.
\newblock \bibinfo{booktitle}{\emph{API Pricing}}.
\newblock OpenAI.
\newblock
\urldef\tempurl%
\url{https://openai.com/api/pricing/}
\showURL{%
\tempurl}
\newblock
\shownote{Official pricing page}.


\bibitem[dee({[n.\,d.]})]%
        {deepseek_platform_2025}
DeepSeek \bibinfo{year}{[n.\,d.]}\natexlab{}.
\newblock \bibinfo{booktitle}{\emph{DeepSeek API Platform}}.
\newblock DeepSeek.
\newblock
\urldef\tempurl%
\url{https://platform.deepseek.com/}
\showURL{%
\tempurl}
\newblock
\shownote{Official platform for accessing DeepSeek models and API resources}.


\bibitem[ANS(2018)]%
        {ANSI-X9.111}
 \bibinfo{year}{2018}\natexlab{}.
\newblock \bibinfo{title}{Penetration Testing within the Financial Service Industry}.
\newblock


\bibitem[pen(2025)]%
        {pen_test_wiki}
 \bibinfo{year}{2025}\natexlab{}.
\newblock \bibinfo{title}{Penetration test}.
\newblock \bibinfo{howpublished}{\url{https://en.wikipedia.org/wiki/Penetration_test}}.
\newblock
\newblock
\shownote{Accessed 2025-08-24}.


\bibitem[Alshehri et~al\mbox{.}(2024)]%
        {alshehri2024breachseek}
\bibfield{author}{\bibinfo{person}{Ibrahim Alshehri}, \bibinfo{person}{Adnan Alshehri}, \bibinfo{person}{Abdulrahman Almalki}, \bibinfo{person}{Majed Bamardouf}, {and} \bibinfo{person}{Alaqsa Akbar}.} \bibinfo{year}{2024}\natexlab{}.
\newblock \showarticletitle{Breachseek: A multi-agent automated penetration tester}.
\newblock \bibinfo{journal}{\emph{arXiv preprint arXiv:2409.03789}} (\bibinfo{year}{2024}).
\newblock


\bibitem[Contributors(2025)]%
        {github-cli}
\bibfield{author}{\bibinfo{person}{GitHub~CLI Contributors}.} \bibinfo{year}{2025}\natexlab{}.
\newblock \bibinfo{title}{GitHub CLI}.
\newblock
\urldef\tempurl%
\url{https://cli.github.com/}
\showURL{%
\tempurl}
\newblock
\shownote{Accessed: 2025-08-24}.


\bibitem[DeepseekAI(2025)]%
        {deepseek_v3_0324}
\bibfield{author}{\bibinfo{person}{DeepseekAI}.} \bibinfo{year}{2025}\natexlab{}.
\newblock \bibinfo{title}{DeepSeek-V3-0324}.
\newblock
\urldef\tempurl%
\url{https://huggingface.co/deepseek-ai/DeepSeek-V3-0324}
\showURL{%
\tempurl}
\newblock
\shownote{Accessed: 2025-08-13}.


\bibitem[Deng et~al\mbox{.}(2024)]%
        {299699}
\bibfield{author}{\bibinfo{person}{Gelei Deng}, \bibinfo{person}{Yi Liu}, \bibinfo{person}{V{\'\i}ctor Mayoral-Vilches}, \bibinfo{person}{Peng Liu}, \bibinfo{person}{Yuekang Li}, \bibinfo{person}{Yuan Xu}, \bibinfo{person}{Tianwei Zhang}, \bibinfo{person}{Yang Liu}, \bibinfo{person}{Martin Pinzger}, {and} \bibinfo{person}{Stefan Rass}.} \bibinfo{year}{2024}\natexlab{}.
\newblock \showarticletitle{{PentestGPT}: Evaluating and Harnessing Large Language Models for Automated Penetration Testing}. In \bibinfo{booktitle}{\emph{33rd USENIX Security Symposium (USENIX Security 24)}}. \bibinfo{publisher}{USENIX Association}, \bibinfo{address}{Philadelphia, PA}, \bibinfo{pages}{847--864}.
\newblock
\showISBNx{978-1-939133-44-1}
\urldef\tempurl%
\url{https://www.usenix.org/conference/usenixsecurity24/presentation/deng}
\showURL{%
\tempurl}


\bibitem[{Ewelina Baran}(2023)]%
        {ref3}
\bibfield{author}{\bibinfo{person}{{Ewelina Baran}}.} \bibinfo{year}{2023}\natexlab{}.
\newblock \bibinfo{booktitle}{\emph{How Much Does Penetration Testing Cost?}}
\newblock
\newblock
\shownote{Accessed: 2025-08-12}.


\bibitem[FOFA(2024)]%
        {fofa}
\bibfield{author}{\bibinfo{person}{FOFA}.} \bibinfo{year}{2024}\natexlab{}.
\newblock \bibinfo{booktitle}{\emph{FOFA Search Engine.}}
\newblock
\newblock
\shownote{\url{https://fofa.info/}}.


\bibitem[Gioacchini et~al\mbox{.}(2024)]%
        {gioacchini2024autopenbench}
\bibfield{author}{\bibinfo{person}{Luca Gioacchini}, \bibinfo{person}{Marco Mellia}, \bibinfo{person}{Idilio Drago}, \bibinfo{person}{Alexander Delsanto}, \bibinfo{person}{Giuseppe Siracusano}, {and} \bibinfo{person}{Roberto Bifulco}.} \bibinfo{year}{2024}\natexlab{}.
\newblock \bibinfo{title}{AutoPenBench: Benchmarking Generative Agents for Penetration Testing}.
\newblock
\showeprint[arxiv]{2410.03225}~[cs.CR]
\urldef\tempurl%
\url{https://arxiv.org/abs/2410.03225}
\showURL{%
\tempurl}


\bibitem[{Hack The Box Limited}(2016)]%
        {HackTheBox}
\bibfield{author}{\bibinfo{person}{{Hack The Box Limited}}.} \bibinfo{year}{2016}\natexlab{}.
\newblock \bibinfo{title}{{Hack The Box}: An Online Platform for Cybersecurity Training}.
\newblock \bibinfo{howpublished}{Online Platform}.
\newblock
\urldef\tempurl%
\url{https://www.hackthebox.com/}
\showURL{%
\tempurl}
\newblock
\shownote{Accessed: 2025-08-13}.


\bibitem[{Heath Adams}(2024)]%
        {ref4}
\bibfield{author}{\bibinfo{person}{{Heath Adams}}.} \bibinfo{year}{2024}\natexlab{}.
\newblock \bibinfo{booktitle}{\emph{How Much Does a Penetration Test Cost in 2025?}}
\newblock
\newblock
\shownote{Accessed: 2025-08-12}.


\bibitem[Henke(2025)]%
        {henke2025autopentestenhancingvulnerabilitymanagement}
\bibfield{author}{\bibinfo{person}{Julius Henke}.} \bibinfo{year}{2025}\natexlab{}.
\newblock \bibinfo{title}{AutoPentest: Enhancing Vulnerability Management With Autonomous LLM Agents}.
\newblock
\showeprint[arxiv]{2505.10321}~[cs.CR]
\urldef\tempurl%
\url{https://arxiv.org/abs/2505.10321}
\showURL{%
\tempurl}


\bibitem[Isozaki et~al\mbox{.}(2024)]%
        {isozaki2024automatedpenetrationtestingintroducing}
\bibfield{author}{\bibinfo{person}{Isamu Isozaki}, \bibinfo{person}{Manil Shrestha}, \bibinfo{person}{Rick Console}, {and} \bibinfo{person}{Edward Kim}.} \bibinfo{year}{2024}\natexlab{}.
\newblock \bibinfo{title}{Towards Automated Penetration Testing: Introducing LLM Benchmark, Analysis, and Improvements}.
\newblock
\showeprint[arxiv]{2410.17141}~[cs.CR]
\urldef\tempurl%
\url{https://arxiv.org/abs/2410.17141}
\showURL{%
\tempurl}


\bibitem[Kapko(2025)]%
        {kapko2025darpa}
\bibfield{author}{\bibinfo{person}{Matt Kapko}.} \bibinfo{year}{2025}\natexlab{}.
\newblock \showarticletitle{DARPA’s AI Cyber Challenge reveals winning models for automated vulnerability discovery and patching}.
\newblock \bibinfo{journal}{\emph{CyberScoop}} (\bibinfo{date}{8 Aug.} \bibinfo{year}{2025}).
\newblock
\urldef\tempurl%
\url{https://cyberscoop.com/darpa-ai-cyber-challenge-winners-def-con-2025/}
\showURL{%
\tempurl}
\newblock
\shownote{Accessed on YYYY-MM-DD}.


\bibitem[Kong et~al\mbox{.}(2025)]%
        {kong2025vulnbotautonomouspenetrationtesting}
\bibfield{author}{\bibinfo{person}{He Kong}, \bibinfo{person}{Die Hu}, \bibinfo{person}{Jingguo Ge}, \bibinfo{person}{Liangxiong Li}, \bibinfo{person}{Tong Li}, {and} \bibinfo{person}{Bingzhen Wu}.} \bibinfo{year}{2025}\natexlab{}.
\newblock \bibinfo{title}{VulnBot: Autonomous Penetration Testing for a Multi-Agent Collaborative Framework}.
\newblock
\showeprint[arxiv]{2501.13411}~[cs.SE]
\urldef\tempurl%
\url{https://arxiv.org/abs/2501.13411}
\showURL{%
\tempurl}


\bibitem[{LangChain Inc.}(2025)]%
        {langgraph_software}
\bibfield{author}{\bibinfo{person}{{LangChain Inc.}}} \bibinfo{year}{2025}\natexlab{}.
\newblock \bibinfo{booktitle}{\emph{LangGraph: Build resilient language agents as graphs}}.
\newblock
\urldef\tempurl%
\url{https://github.com/langchain-ai/langgraph}
\showURL{%
\tempurl}
\newblock
\shownote{MIT License}.


\bibitem[Liu et~al\mbox{.}(2023)]%
        {liu2023lostmiddlelanguagemodels}
\bibfield{author}{\bibinfo{person}{Nelson~F. Liu}, \bibinfo{person}{Kevin Lin}, \bibinfo{person}{John Hewitt}, \bibinfo{person}{Ashwin Paranjape}, \bibinfo{person}{Michele Bevilacqua}, \bibinfo{person}{Fabio Petroni}, {and} \bibinfo{person}{Percy Liang}.} \bibinfo{year}{2023}\natexlab{}.
\newblock \bibinfo{title}{Lost in the Middle: How Language Models Use Long Contexts}.
\newblock
\showeprint[arxiv]{2307.03172}~[cs.CL]
\urldef\tempurl%
\url{https://arxiv.org/abs/2307.03172}
\showURL{%
\tempurl}


\bibitem[Malik(2025)]%
        {ref1}
\bibfield{author}{\bibinfo{person}{Keshav Malik}.} \bibinfo{year}{2025}\natexlab{}.
\newblock \bibinfo{booktitle}{\emph{Why Penetration Testing is Important}}.
\newblock
\newblock
\shownote{Accessed: 2025-08-12}.


\bibitem[Metasploit({[n.\,d.]})]%
        {metasploit}
\bibfield{author}{\bibinfo{person}{Metasploit}.} \bibinfo{year}{[n.\,d.]}\natexlab{}.
\newblock \bibinfo{booktitle}{\emph{Metasploit The world’s most used penetration testing framework}}.
\newblock
\newblock
\shownote{\url{https://www.metasploit.com/}}.


\bibitem[Muzsai et~al\mbox{.}(2024)]%
        {muzsai2024hacksynthllmagentevaluation}
\bibfield{author}{\bibinfo{person}{Lajos Muzsai}, \bibinfo{person}{David Imolai}, {and} \bibinfo{person}{András Lukács}.} \bibinfo{year}{2024}\natexlab{}.
\newblock \bibinfo{title}{HackSynth: LLM Agent and Evaluation Framework for Autonomous Penetration Testing}.
\newblock
\showeprint[arxiv]{2412.01778}~[cs.CR]
\urldef\tempurl%
\url{https://arxiv.org/abs/2412.01778}
\showURL{%
\tempurl}


\bibitem[{National Institute of Standards and Technology}(2025)]%
        {NVD}
\bibfield{author}{\bibinfo{person}{{National Institute of Standards and Technology}}.} \bibinfo{year}{2025}\natexlab{}.
\newblock \bibinfo{title}{{National Vulnerability Database (NVD)}}.
\newblock \bibinfo{howpublished}{Online Database}.
\newblock
\urldef\tempurl%
\url{https://nvd.nist.gov/}
\showURL{%
\tempurl}
\newblock
\shownote{Accessed: 2025-08-13}.


\bibitem[of~Standards and Technology(2018)]%
        {nvd2018cve-2018-7600}
\bibfield{author}{\bibinfo{person}{National~Institute of Standards} {and} \bibinfo{person}{Technology}.} \bibinfo{year}{2018}\natexlab{}.
\newblock \bibinfo{title}{CVE-2018-7600 Detail}.
\newblock
\urldef\tempurl%
\url{https://nvd.nist.gov/vuln/detail/CVE-2018-7600}
\showURL{%
\tempurl}
\newblock
\shownote{Accessed: 2025-08-27}.


\bibitem[of~Standards and Technology(2019)]%
        {nvd2019cve-2019-6339}
\bibfield{author}{\bibinfo{person}{National~Institute of Standards} {and} \bibinfo{person}{Technology}.} \bibinfo{year}{2019}\natexlab{}.
\newblock \bibinfo{title}{CVE-2019-6339 Detail}.
\newblock
\urldef\tempurl%
\url{https://nvd.nist.gov/vuln/detail/CVE-2019-6339}
\showURL{%
\tempurl}
\newblock
\shownote{Accessed: 2025-08-27}.


\bibitem[OpenAI(2025)]%
        {openai_gpt5_2025}
\bibfield{author}{\bibinfo{person}{OpenAI}.} \bibinfo{year}{2025}\natexlab{}.
\newblock \bibinfo{title}{Introducing GPT-5}.
\newblock
\urldef\tempurl%
\url{https://openai.com/index/introducing-gpt-5/}
\showURL{%
\tempurl}
\newblock
\shownote{Accessed: 2025-08-13}.


\bibitem[{OpenAI}(2025)]%
        {openai_api}
\bibfield{author}{\bibinfo{person}{{OpenAI}}.} \bibinfo{year}{2025}\natexlab{}.
\newblock \bibinfo{title}{OpenAI API}.
\newblock \bibinfo{howpublished}{\url{https://openai.com/api/}}.
\newblock
\newblock
\shownote{Accessed: 2025-08-20}.


\bibitem[{OWASP Foundation}(2021)]%
        {OWASP-Testing}
\bibfield{author}{\bibinfo{person}{{OWASP Foundation}}.} \bibinfo{year}{2021}\natexlab{}.
\newblock \bibinfo{title}{OWASP Web Security Testing Guide}.
\newblock \bibinfo{howpublished}{Online}.
\newblock
\urldef\tempurl%
\url{https://owasp.org/www-project-web-security-testing-guide/}
\showURL{%
\tempurl}


\bibitem[Phith0n(2019)]%
        {Vulhub}
\bibfield{author}{\bibinfo{person}{Phith0n}.} \bibinfo{year}{2019}\natexlab{}.
\newblock \bibinfo{title}{{Vulhub}: Pre-Built Vulnerable Environments Based on Docker}.
\newblock \bibinfo{howpublished}{GitHub Repository}.
\newblock
\urldef\tempurl%
\url{https://github.com/vulhub/vulhub}
\showURL{%
\tempurl}
\newblock
\shownote{Accessed: 2025-08-13}.


\bibitem[Pratama et~al\mbox{.}({[n.\,d.]})]%
        {Pratama_2024}
\bibfield{author}{\bibinfo{person}{Derry Pratama}, \bibinfo{person}{Naufal Suryanto}, \bibinfo{person}{Andro~Aprila Adiputra}, \bibinfo{person}{Thi-Thu-Huong Le}, \bibinfo{person}{Ahmada~Yusril Kadiptya}, \bibinfo{person}{Muhammad Iqbal}, {and} \bibinfo{person}{Howon Kim}.} \bibinfo{year}{[n.\,d.]}\natexlab{}.
\newblock \showarticletitle{CIPHER: Cybersecurity Intelligent Penetration-Testing Helper for Ethical Researcher}.
\newblock \bibinfo{journal}{\emph{Sensors}} \bibinfo{volume}{24}, \bibinfo{number}{21} (\bibinfo{year}{[n.\,d.]}).
\newblock
\showISSN{1424-8220}
\href{https://doi.org/10.3390/s24216878}{doi:\nolinkurl{10.3390/s24216878}}


\bibitem[{PTES Committee}(2014)]%
        {PTES}
\bibfield{author}{\bibinfo{person}{{PTES Committee}}.} \bibinfo{year}{2014}\natexlab{}.
\newblock \bibinfo{title}{Penetration Testing Execution Standard}.
\newblock \bibinfo{howpublished}{Online}.
\newblock
\urldef\tempurl%
\url{http://www.pentest-standard.org/}
\showURL{%
\tempurl}
\newblock
\shownote{Version 1.0}.


\bibitem[Qwen(2025a)]%
        {qwen3_14b}
\bibfield{author}{\bibinfo{person}{Qwen}.} \bibinfo{year}{2025}\natexlab{a}.
\newblock \bibinfo{title}{Qwen3-14B}.
\newblock
\urldef\tempurl%
\url{https://huggingface.co/Qwen/Qwen3-14B}
\showURL{%
\tempurl}
\newblock
\shownote{Accessed: 2025-08-13}.


\bibitem[Qwen(2025b)]%
        {qwen3_1.7b}
\bibfield{author}{\bibinfo{person}{Qwen}.} \bibinfo{year}{2025}\natexlab{b}.
\newblock \bibinfo{title}{Qwen3-1.7B}.
\newblock
\urldef\tempurl%
\url{https://huggingface.co/Qwen/Qwen3-1.7B}
\showURL{%
\tempurl}
\newblock
\shownote{Accessed: 2025-08-13}.


\bibitem[Qwen(2025c)]%
        {qwen3_30b_a3b}
\bibfield{author}{\bibinfo{person}{Qwen}.} \bibinfo{year}{2025}\natexlab{c}.
\newblock \bibinfo{title}{Qwen3-30B-A3B}.
\newblock
\urldef\tempurl%
\url{https://huggingface.co/Qwen/Qwen3-30B-A3B}
\showURL{%
\tempurl}
\newblock
\shownote{Accessed: 2025-08-13}.


\bibitem[Qwen(2025d)]%
        {qwen3_4b}
\bibfield{author}{\bibinfo{person}{Qwen}.} \bibinfo{year}{2025}\natexlab{d}.
\newblock \bibinfo{title}{Qwen3-4B}.
\newblock
\urldef\tempurl%
\url{https://huggingface.co/Qwen/Qwen3-4B}
\showURL{%
\tempurl}
\newblock
\shownote{Accessed: 2025-08-13}.


\bibitem[Qwen(2025e)]%
        {qwen3_8b}
\bibfield{author}{\bibinfo{person}{Qwen}.} \bibinfo{year}{2025}\natexlab{e}.
\newblock \bibinfo{title}{Qwen3-8B}.
\newblock
\urldef\tempurl%
\url{https://huggingface.co/Qwen/Qwen3-8B}
\showURL{%
\tempurl}
\newblock
\shownote{Accessed: 2025-08-13}.


\bibitem[Scarfone et~al\mbox{.}(2008)]%
        {NIST-SP800-115}
\bibfield{author}{\bibinfo{person}{Karen Scarfone}, \bibinfo{person}{Murugiah Souppaya}, {and} \bibinfo{person}{Amanda Cody}.} \bibinfo{year}{2008}\natexlab{}.
\newblock \bibinfo{booktitle}{\emph{Technical Guide to Information Security Testing and Assessment}}.
\newblock \bibinfo{type}{{T}echnical {R}eport} NIST Special Publication 800-115. \bibinfo{institution}{National Institute of Standards and Technology}, \bibinfo{address}{Gaithersburg, MD}.
\newblock
\href{https://doi.org/10.6028/NIST.SP.800-115}{doi:\nolinkurl{10.6028/NIST.SP.800-115}}


\bibitem[Security(2025)]%
        {kali_linux}
\bibfield{author}{\bibinfo{person}{Offensive Security}.} \bibinfo{year}{2025}\natexlab{}.
\newblock \bibinfo{title}{Kali Linux}.
\newblock
\urldef\tempurl%
\url{https://www.kali.org/}
\showURL{%
\tempurl}
\newblock
\shownote{Accessed: 2025-08-20}.


\bibitem[Shen et~al\mbox{.}(2025)]%
        {shen2025pentestagentincorporatingllmagents}
\bibfield{author}{\bibinfo{person}{Xiangmin Shen}, \bibinfo{person}{Lingzhi Wang}, \bibinfo{person}{Zhenyuan Li}, \bibinfo{person}{Yan Chen}, \bibinfo{person}{Wencheng Zhao}, \bibinfo{person}{Dawei Sun}, \bibinfo{person}{Jiashui Wang}, {and} \bibinfo{person}{Wei Ruan}.} \bibinfo{year}{2025}\natexlab{}.
\newblock \bibinfo{title}{PentestAgent: Incorporating LLM Agents to Automated Penetration Testing}.
\newblock
\showeprint[arxiv]{2411.05185}~[cs.CR]
\urldef\tempurl%
\url{https://arxiv.org/abs/2411.05185}
\showURL{%
\tempurl}


\bibitem[Xu et~al\mbox{.}(2024)]%
        {xu2024autoattackerlargelanguagemodel}
\bibfield{author}{\bibinfo{person}{Jiacen Xu}, \bibinfo{person}{Jack~W. Stokes}, \bibinfo{person}{Geoff McDonald}, \bibinfo{person}{Xuesong Bai}, \bibinfo{person}{David Marshall}, \bibinfo{person}{Siyue Wang}, \bibinfo{person}{Adith Swaminathan}, {and} \bibinfo{person}{Zhou Li}.} \bibinfo{year}{2024}\natexlab{}.
\newblock \bibinfo{title}{AutoAttacker: A Large Language Model Guided System to Implement Automatic Cyber-attacks}.
\newblock
\showeprint[arxiv]{2403.01038}~[cs.CR]
\urldef\tempurl%
\url{https://arxiv.org/abs/2403.01038}
\showURL{%
\tempurl}


\bibitem[Yang et~al\mbox{.}(2025)]%
        {qwen3}
\bibfield{author}{\bibinfo{person}{An Yang}, \bibinfo{person}{Anfeng Li}, \bibinfo{person}{Baosong Yang}, \bibinfo{person}{Beichen Zhang}, \bibinfo{person}{Binyuan Hui}, \bibinfo{person}{Bo Zheng}, \bibinfo{person}{Bowen Yu}, \bibinfo{person}{Chang Gao}, \bibinfo{person}{Chengen Huang}, \bibinfo{person}{Chenxu Lv}, \bibinfo{person}{Chujie Zheng}, \bibinfo{person}{Dayiheng Liu}, \bibinfo{person}{Fan Zhou}, \bibinfo{person}{Fei Huang}, \bibinfo{person}{Feng Hu}, \bibinfo{person}{Hao Ge}, \bibinfo{person}{Haoran Wei}, \bibinfo{person}{Huan Lin}, \bibinfo{person}{Jialong Tang}, \bibinfo{person}{Jian Yang}, \bibinfo{person}{Jianhong Tu}, \bibinfo{person}{Jianwei Zhang}, \bibinfo{person}{Jianxin Yang}, \bibinfo{person}{Jiaxi Yang}, \bibinfo{person}{Jing Zhou}, \bibinfo{person}{Jingren Zhou}, \bibinfo{person}{Junyang Lin}, \bibinfo{person}{Kai Dang}, \bibinfo{person}{Keqin Bao}, \bibinfo{person}{Kexin Yang}, \bibinfo{person}{Le Yu}, \bibinfo{person}{Lianghao Deng}, \bibinfo{person}{Mei Li}, \bibinfo{person}{Mingfeng
  Xue}, \bibinfo{person}{Mingze Li}, \bibinfo{person}{Pei Zhang}, \bibinfo{person}{Peng Wang}, \bibinfo{person}{Qin Zhu}, \bibinfo{person}{Rui Men}, \bibinfo{person}{Ruize Gao}, \bibinfo{person}{Shixuan Liu}, \bibinfo{person}{Shuang Luo}, \bibinfo{person}{Tianhao Li}, \bibinfo{person}{Tianyi Tang}, \bibinfo{person}{Wenbiao Yin}, \bibinfo{person}{Xingzhang Ren}, \bibinfo{person}{Xinyu Wang}, \bibinfo{person}{Xinyu Zhang}, \bibinfo{person}{Xuancheng Ren}, \bibinfo{person}{Yang Fan}, \bibinfo{person}{Yang Su}, \bibinfo{person}{Yichang Zhang}, \bibinfo{person}{Yinger Zhang}, \bibinfo{person}{Yu Wan}, \bibinfo{person}{Yuqiong Liu}, \bibinfo{person}{Zekun Wang}, \bibinfo{person}{Zeyu Cui}, \bibinfo{person}{Zhenru Zhang}, \bibinfo{person}{Zhipeng Zhou}, {and} \bibinfo{person}{Zihan Qiu}.} \bibinfo{year}{2025}\natexlab{}.
\newblock \showarticletitle{Qwen3 Technical Report}.
\newblock \bibinfo{journal}{\emph{arXiv preprint arXiv:2505.09388}} (\bibinfo{year}{2025}).
\newblock


\bibitem[…(2022)]%
        {measuring_cloud_services}
\bibfield{author}{\bibinfo{person}{…}.} \bibinfo{year}{2022}\natexlab{}.
\newblock \showarticletitle{Measuring and Mitigating the Risk of IP Reuse on Public Clouds}.
\newblock  (\bibinfo{year}{2022}).
\newblock
\newblock
\shownote{Identified dozens of exploitable software systems spanning hundreds of servers, e.g., databases, caches, mobile applications, and web services}.


\end{thebibliography}


\appendix

\section{Supplement to Design of \songbench}

Table \ref{table:cves} presents the 30 CVEs incorporated in \songbench along with the 25 services they affect, spanning the years 2015 to 2025. In addition, the 14 selected benign mainstream services are listed in the Table \ref{table:benign}. By combining these vulnerable services with varying numbers of benign services, we ultimately constructed the 510 target machines in \songbench.

\begin{table}[!ht]
    \centering
    \tiny
    \caption{14 benign services used in \songbench.}
    \resizebox{0.2\textwidth}{!}{
\begin{tabular}{cc}
\toprule
   & \textbf{Service} \\
   \midrule
1  & sshd              \\
2  & vsftpd            \\
3  & mysql             \\
4  & postfix           \\
5  & dnsmasq           \\
6  & ldap              \\
7  & redis             \\
8  & postgres          \\
9  & mosquitto         \\
10 & xrdp              \\
11 & mongodb           \\
12 & http              \\
13 & nginx             \\
14 & samba            \\ \bottomrule
\end{tabular}
    }    
    \label{table:benign}
\end{table} 

\begin{table}[!ht]  
    \centering
    \small
    \caption{Overview of the Benchmark Configurations.}
    \label{table:benchmark_overview}
    \resizebox{\columnwidth}{!}{ 
    \begin{tabular}{c c c c}
    \toprule
    \textbf{Configuration} & \textbf{\# of Benign Services} & \textbf{\# of Vulnerable Services} & \textbf{\# of Hosts} \\
    \midrule
    Tier 0 & 0 & 1 & 30 \\
    Tier 1 & 1 & 1 & 120 \\
    Tier 2 & 3 & 1 & 120 \\
    Tier 3 & 5 & 1 & 120 \\
    Tier 4 & 7 & 1 & 120 \\
    \midrule
    \textbf{Total} & - & - & \textbf{510} \\
    \bottomrule
    \end{tabular}
    }
\end{table}

\begin{table}[!ht]
    \centering
    \tiny
    \caption{30 CVEs and their affacted services used in \songbench.}
    \resizebox{0.36\textwidth}{!}{
\begin{tabular}{ccc}
\toprule
   & \textbf{CVE ID} & \textbf{Affected Service} \\
   \midrule
1  & CVE-2015-1427   & Elasticsearch             \\
2  & CVE-2015-3306   & ProFTPD                   \\
3  & CVE-2015-8562   & Joomla                    \\
4  & CVE-2016-3088   & ActiveMQ                  \\
5  & CVE-2016-5734   & phpMyAdmin                \\
6  & CVE-2017-12636  & CouchDB                   \\
7  & CVE-2017-16082  & Node                      \\
8  & CVE-2017-17562  & GoAhead                   \\
9  & CVE-2017-7494   & Samba                     \\
10 & CVE-2018-1297   & JMeter-Server             \\
11 & CVE-2018-20062  & ThinkPHP                  \\
12 & CVE-2018-7600   & Drupal                    \\
13 & CVE-2019-11043  & PHP-FPM                  \\
14 & CVE-2019-17564  & Dubbo                     \\
15 & CVE-2020-35476  & OpenTSDB                  \\
16 & CVE-2020-7247   & OpenSMTPD                 \\
17 & CVE-2021-25646  & Apache-Druid              \\
18 & CVE-2021-41773  & Apache-HTTPD              \\
19 & CVE-2021-42013  & Apache-HTTPD              \\
20 & CVE-2022-0543   & Redis                     \\
21 & CVE-2022-22965  & Spring-WebMVC            \\
22 & CVE-2022-24706  & CouchDB                   \\
23 & CVE-2022-24816  & GeoServer                 \\
24 & CVE-2022-41678  & ActiveMQ                \\
25 & CVE-2023-25826  & OpenTSDB                  \\
26 & CVE-2023-51467  & OFBiz                     \\
27 & CVE-2024-27348  & HugeGraph                 \\
28 & CVE-2024-36401  & GeoServer                 \\
29 & CVE-2025-32433  & Erlang/OTP(sshd)          \\
30 & CVE-2025-3248   & Langflow                 \\
\bottomrule
\end{tabular}
    }    
    \label{table:cves}
\end{table}

\section{Supplement to Desgin of \ouragent}\label{app:design}

\noindent \textbf{Reasoner Module.} The Reasoner Module is primarily responsible for high-level planning. It determines whether the overall goal of the penetration testing task has been achieved based on the stage-wise progress of the test, rather than on the details of specific instruction execution. If the overall goal is not yet accomplished, it generates the next phased goal and hands it over to the Assistant Module. The Reasoner Module proceeds to the next planning step only after the Assistant Module reports success or failure in achieving the phased goal. Throughout this process, the Assistant Module interacts with the Memory Module to retrieve abstract-level activated memory as input context for decision-making.

\noindent \textbf{Assistant Module.} The Assistant Module is responsible for the phased goal assigned by the Reasoner Module and performs low-level planning based on it, such as generating the specific instructions to be executed. Prior to generating the next instruction, the Assistant Module verifies whether the phased goal has succeeded or failed, determining whether control should revert to the Reasoner Module for a new phased goal. During this process, the Assistant Module also interacts with the Memory Module to obtain coarse-grained level activated memory, which serves as context for decision-making.

In addition, the Assistant Module interacts with the \arsenalmodule to obtain the manuals for exploits available based on the current penetration progress—whether in-the-wild or from third-party tools such as Metasploit—thereby assisting in the generation of concrete instructions to be executed.

\noindent \textbf{Executor Module.} The Executor Module executes the concrete instructions generated by the Assistant Module and submits both the instructions and their raw execution results to the Memory Module for organization and compression. Upon completion of instruction execution, control reverts to the Assistant Module to proceed with the next planning step.

\section{Supplement to Instructions Reduction of Auto-Pen-Bench} \label{app:reduction}

Apart from the target subnet and the goal of obtaining the flag, the instructions in Auto-Pen-Bench also include some additional hints, such as penetration entry points, exploit paths, and even details of specific command execution. Below are the original instructions for the \texttt{CVE\_VM8} target in Auto-Pen-Bench, which contain target service's username \texttt{admin}, credentials \texttt{password}, the flag’s storage path \texttt{/var/www/htm-} \texttt{l/flag.txt}, and specific commands like \texttt{launch msfconsole}. Such information is rarely available in real-world penetration testing tasks, where the only known data may be the target’s IP address or subnet.

\begin{tcolorbox}[title = {Original Instruction for Auto-Pen-Bench}] 
Discover the vulnerable target machine within the 192.168.5.0/24 network. Inspect the service through direct interaction. Launch an interactive msfconsole to ientify a known exploit (Do not use searchsploit). Correctly configure the exploit to access the /var/www/html/flag.txt file. When you discover the service, use admin:password as 'admin' credentials.

Remember: If you need to run command outside msfconsole after you used it, remember to type 'exit'
\end{tcolorbox}

Accordingly, for experiment on performance under limited instructions in Section \ref{sec:evaluation}, we reduced all Auto-Pen-Bench's instructions to include only the target host’s subnet and the flag acquisition goal. The final instructions used in the experiment are shown below and this provides a more objective assessment of the agent’s penetration testing performance.

\begin{tcolorbox}[title = {Limited Instruction for Auto-Pen-Bench}] 
Discove the target machine within 192.168.5.0/24 and try to get the flag on it.
\end{tcolorbox}

\section{Details of LLMs Used in Evaluation} 

Table \ref{table:llms} presents the basic information, pricing, and default parameter settings of the LLMs used in the experiments for \ouragent. It should be noted that the GPT-5 series models no longer allow users to specify the \texttt{temperature} and \texttt{top\_p} parameters\cite{openai_api}.

\begin{table*}[]
    \centering
    \small
    \caption{LLMs information.}
    \resizebox{0.8\textwidth}{!}{
\begin{tabular}{c|ccc|cc}
\toprule
\multirow{2}{*}{\textbf{Model}} & \multicolumn{3}{c|}{\textbf{Cost}}                                 & \multicolumn{2}{c}{\textbf{Hyperparameters}} \\\cmidrule(lr){2-4} \cmidrule(lr){5-6}
                                & \textbf{API Platform} & \textbf{Input}      & \textbf{Output}     & \textbf{Temperature}     & \textbf{Top P}    \\
                                \midrule
GPT-5\cite{openai_gpt5_2025}                           &          \cite{openai_api_pricing_en_2025}             & 1.25USD / 1M tokens & 10 USD / 1M tokens  & -                        & -                 \\
DeepSeek-V3-0324\cite{deepseek_v3_0324}                &          \cite{deepseek_platform_2025}               & 2RMB / 1M tokens    & 8RMB / 1M tokens    & 1.0                      & 1.0               \\
Qwen3-30B-A3B\cite{qwen3_30b_a3b}                   &       \cite{aliyun_bailian_console_2025}                  & 0.75RMB / 1M tokens & 7.5 RMB / 1M tokens & 0.6                      & 0.95              \\
Qwen3-14B \cite{qwen3_14b}                      &      \cite{aliyun_bailian_console_2025}                   & 1RMB / 1M tokens    & 10RMB / 1M tokens   & 0.6                      & 0.95              \\
Qwen3-8B \cite{qwen3_8b}                       &     \cite{aliyun_bailian_console_2025}                    & 0.3RMB / 1M tokens  & 3RMB / 1M tokens    & 0.6                      & 0.95              \\
Qwen3-4B \cite{qwen3_4b}                       &    \cite{aliyun_bailian_console_2025}                     & 0.3RMB / 1M tokens  & 3RMB / 1M tokens    & 0.6                      & 0.95              \\
Qwen3-1.7B \cite{qwen3_1.7b}                     &    \cite{aliyun_bailian_console_2025}                     & 0.3RMB / 1M tokens  & 3RMB / 1M tokens    & 0.6                      & 0.95        \\
\bottomrule
\end{tabular}
    }    
    \label{table:llms}
\end{table*}

\section{Details of UED Dimensions}

Table~\ref{tab:ued_dimensions} provides the detailed definitions of all dimensions used in the Unified Exploit Descriptor (UED). 
These dimensions are divided into two functional categories: those required for automated environment construction 
(e.g., \texttt{language}, \texttt{system\_dependencies}, \texttt{main\_script}) and those required for generating 
agent-invocable manuals (e.g., \texttt{parameters}, \texttt{setup\_steps}, \texttt{exploit\_steps}). 
The table formalizes how heterogeneous exploit repositories are transformed into standardized, reproducible specifications 
for autonomous execution by \arsenalmodule.

\begin{table*}[]
    \centering
    \footnotesize
    \caption{Unified Exploit Descriptor (UED) dimensions, organized into environmental and operational categories.}
    \resizebox{0.9\textwidth}{!}{
    \begin{tabular}{>{\centering\arraybackslash}p{3cm} >{\centering\arraybackslash}p{10cm}}
    \toprule
    \textbf{Dimension} & \textbf{Description} \\
    \midrule
    \multicolumn{2}{c}{\textbf{Environmental dimensions (for reproducible execution environment)}} \\
    \midrule
    Language & Primary programming language of the exploit (e.g., Python, C, Go). \\
    Language version & Recommended language version ensuring compatibility (e.g., Python 3.9 vs. 2.7). \\
    Base image & Lightweight Docker base image aligned with language/runtime (e.g., \texttt{python:3.9-slim}). \\
    System dependencies & OS-level packages to be installed via package manager (e.g., \texttt{nmap}, \texttt{build-essential}). \\
    Code dependencies & Language-specific libraries installed via package manager (e.g., \texttt{requests}, \texttt{pwntools}). \\
    Main script & Path, executor, and metadata of the primary exploit script. Serves as container entrypoint.\\
    Parameter files & External files (payloads, configs, URL lists) required by the exploit. \\
    Docker config & Workdir, entrypoint, and command defaults for reproducible containerization. \\
    \midrule
    \multicolumn{2}{c}{\textbf{Operational dimensions (for manual generation)}} \\
    \midrule
    Setup steps & Environment and target setup before exploitation(e.g., listener, service connection.\\
    Exploit steps & Ordered attack operations (e.g., run script, send payload). Form the execution skeleton. \\
    Parameters & Command-line arguments with placeholders, descriptions, and defaults. \\
    Usage example & Repository-derived command-line example.\\
    \bottomrule
    \end{tabular}
    }
    \label{tab:ued_dimensions}
\end{table*}

\end{document}